\documentclass[5p]{elsarticle}

\providecommand{\srcdir}{source}
\providecommand{\refdir}{related}

\usepackage{graphicx}
\usepackage{amsmath,amssymb}
\usepackage{url}
\usepackage{ascmac}
\usepackage{threeparttable}
\usepackage{times}
\usepackage{subfigure}
\usepackage{profile}
\usepackage{slashbox}
\usepackage{multirow}
\usepackage[ruled]{algorithm}
\usepackage{algpseudocode}
\usepackage{caption}

\newcommand{\oc}{$O(1)$}
\newcommand{\ol}{$O(\log{n})$}
\newcommand{\oN}{$O(n)$}
\newcommand{\oM}{$O(m)$}
\newcommand{\oNl}{$O(n\log{n})$}

\newcommand{\lrudll}{LRU$_{DLL}$}
\newcommand{\lrus}{LRU$_{S}$}
\newcommand{\lruc}{LRU$_{C}$}
\newcommand{\lfuh}{LFU$_{H}$}
\newcommand{\cfr}[1][q]{${\rm{}CFR}(#1 )$}

\newcommand{\acsY}[1][]{$A^{#1}_{\rm Real}$}
\newcommand{\acsZ}[2][\alpha ]{$A^{#2 }_{{\rm Zipf}(#1 )}$}
\newcommand{\acsYC}{\acsY[C]}

\newcommand{\acsYPa}{\acsY[P(1.5{\rm KB})]}
\newcommand{\acsYPb}{\acsY[P(15{\rm KB})]}
\newcommand{\acsYPc}{\acsY[P(60{\rm KB})]}
\newcommand{\acsZC}{\acsZ[\alpha ]{C}}

\newcommand{\acsZaC}{\acsZ[0.6]{C}}
\newcommand{\acsZbC}{\acsZ[0.8]{C}}
\newcommand{\acsZcC}{\acsZ[1.0]{C}}
\newcommand{\acsZdC}{\acsZ[1.2]{C}}

\newcommand{\acsZcPb}{\acsZ[1.0]{P(15{\rm KB})}}
\newcommand{\acsZcPc}{\acsZ[1.0]{P(60{\rm KB})}}

\newcommand{\acsZdPc}{\acsZ[1.2]{P(60{\rm KB})}}

\newcommand{\subgraphcachehitrate}[4][0.50]{%
\subfigure[#3]{\includegraphics[clip, width=#1\columnwidth ]%
{\srcdir/CompactCAR/CacheHitRates/006/cache_hit_rate_#2.pdf}%
\label{cache:hit:rate:#4}}%
}

\newcommand{\subgraphcachehitratelinear}[5][0.50]{%
\subfigure[#4 $(c=10^#3)$]{\includegraphics[clip, width=#1\columnwidth ]%
{\srcdir/CompactCAR/CacheHitRates/006/cache_count_hit_prop_#2_10_#3.pdf}%
\label{cache:hit:rate:#5}}%
}

\newcommand{\subgraphreusedistance}[4][0.66]{%
\subfigure[#3]{\includegraphics[clip, width=#1\columnwidth ]%
{\srcdir/CompactCAR/ReuseDistance/02/acs_#2.png}%
\label{reuse:dist:#4}}%
}

\newcommand{\subgraphimprovementratio}[4][0.66]{%
\subfigure[#3]{\includegraphics[clip, width=#1\columnwidth ]%
{\srcdir/CompactCAR/ImprovementRatio/#2.pdf}%
\label{cache:hit:rate:improve:#4}}%
}

\begin{document}

\begin{frontmatter}

\title{Compact CAR: Low-Overhead Cache Replacement Policy for an ICN Router}

\author[aff1]{Atsushi Ooka}
\ead{a-ooka@ist.osaka-u.ac.jp}

\author[aff1]{Suyong Eum}
\ead{suyong@ist.osaka-u.ac.jp}

\author[aff2]{Shingo Ata}
\ead{ata@info.eng.osaka-cu.ac.jp}

\author[aff1]{Masayuki Murata}
\ead{murata@ist.osaka-u.ac.jp}

\address[aff1]{%
Graduate School of Information Science and Technology, Osaka University, \\
1-5 Yamadaoka, Suita, Osaka, 565-0871 Japan \\
Tel.: +81-6-6879-4542, Fax: +81-6-6879-4544}
\address[aff2]{%
Graduate School of Engineering, Osaka City University, \\
3-3-138 Sugimoto, Sumiyoshi-ku, Osaka-shi, Osaka 558-8585, Japan \\
Tel.: +81-6-6605-2191, Fax: +81-6-6690-5382}

%%%%%%%%%%%%%%%%%%%%%%%%%%%%%%%%%%%%%%%%%%%%%%%%%%%%%%%%%%%%%%%%%%%%%%%%
\begin{abstract}

Information-centric networking (ICN) has gained attention from network research communities due to its capability of efficient content dissemination. In-network caching function in ICN plays an important role to achieve the design motivation. However, many researchers on in-network caching have focused on where to cache rather than how to cache: the former is known as contents deployment in the network and the latter is known as cache replacement in an ICN element. Although, the cache replacement has been intensively researched in the context of web-caching and content delivery network previously, the conventional approaches cannot be directly applied to ICN due to the fine granularity of cacheable items in ICN, which eventually changes the access patterns.

In this paper, we argue that ICN requires a novel cache replacement algorithm to fulfill the requirements in the design of high performance ICN element. Then, we propose a novel cache replacement algorithm to satisfy the requirements named Compact CLOCK with Adaptive Replacement (Compact CAR), which can reduce the consumption of cache memory to one-tenth compared to conventional approaches.

\end{abstract}


\end{frontmatter}

%%%%%%%%%%%%%%%%%%%%%%%%%%%%%%%%%%%%%%%%%%%%%%%%%%%%%%%%%%%%%%%%%%%%%%%%%%%%%%%%%%%%%%%%%%%%%%%%%%%%%%%%%%%%%%
\section{Introduction}\label{sed:intro}

Information-centric networking (ICN) was introduced as a future network architecture which is optimized for content dissemination. ICN is built on the idea of name-based routing which enables each ICN element to be aware of users' requests as well as their counterpart responses. Thus, individual ICN elements can be turned into caching devices by simply providing physical cache memory for them.

This feature of ICN that all elements have caching capability is called in-network caching function, and several ICN architectures, CCNx\cite{ccn000}, NDN\cite{ccn003}, SAIL\cite{sail} and PURSUIT\cite{pursuit}, have already suggested utilizing the function to take several advantages of caching system such as reducing network access latency, alleviating network traffic, balancing network load, and achieving robustness against a single failure scenario. In this sense, ICN can be considered as a largely distributed caching architecture whose performance depends on mainly two factors: where to cache and how to cache contents. The former and the latter are known as content placement and cache replacement problems, respectively.

While the problem of content placement has attracted much attention in ICN research communities, that of cache replacement has been ignored since many people believe that the problem has already been investigated intensively in the context of web-caching and content delivery network. However, it is unclear that the conventional cache replacement approaches are suitable for ICN due to following two reasons. First, core ICN elements are expected to meet the speeds required for line-rate operation, especially by exploiting limited memory and computational resources. However, the conventional cache replacement approaches are designed for end-device operation rather than for core-device operation which should be carried out in parallel with forwarding operation. Second, the fine granularity of cacheable items in ICN, namely chunks or segments, changes the traffic access patterns of request messages, which dramatically govern the performance of cache replacement algorithm.

In the light of the observation above, this paper studies the cache replacement problem in the core ICN elements, and so they can support the line-rate operation, which is critical in the design of core ICN elements. First, we analyze the access patterns of contents to understand its relation to cache replacement algorithms. Then, we propose a novel cache replacement algorithm named Compact CLOCK with Adaptive Replacement (Compact CAR) to fulfill the requirements. The proposed algorithm is based on CLOCK that is a classical cache replacement policy to achieve low-complexity approximation. The numerical simulation shows that the proposed cache replacement algorithm can reduce the consumption of cache memory to one-tenth compared to conventional approaches.

This paper is organized as follows. In Section \ref{sec:related}, we review related research works. In Section \ref{sec:problem}, we describe the design considerations of cache replacement algorithm for a core element of ICN.
This is followed by a detail description of our proposed method Compact CAR in Section \ref{sec:proposal}. In Section \ref{sec:eval}, we evaluate our protocol through extensive simulations. Then, we discuss on some
implementation issues of our proposal, especially for the design of high performance of ICN core element in Section \ref{sec:discuss}. Finally, we conclude this article in Section \ref{sec:conclusion}.

%%%%%%%%%%%%%%%%%%%%%%%%%%%%%%%%%%%%%%%%%%%%%%%%%%%%%%%%%%%%%%%%%%%%%%%%%%%%%%%%%%%%%%%%%%%%%%%%%%%%%%%%%%%%%%
\section{Related works}\label{sec:related}

There are a considerable number of cache replacement algorithms, ranging from those available in a computer system (e.g., CPU and I/O buffers) to those used in communication networks (e.g., web-proxies and CDNs). Thus, there are various requirements and methods suitable for the individual environments. To understand the requirements of in-network caching in ICN, we review several cache replacement algorithms that have been carried out in the different context.

Replacement algorithms are developed originally for the purpose of paging in the computer system~\cite{ccn102,ccn108}. The bottleneck of the systems is the latency of fetching pages from slow auxiliary memory to fast cache memory. On the one hand, the hardware cache such as CPU commonly used First-in, first-out (FIFO) and Not Recently Used (NRU) to reduce the cost because of the hardly limited resources. On the other hand, the software cache such as virtual memory of OS commonly adopts LRU and LFU, which are costly to maintain a data structure or/and statistical information (i.e., the number of references to a page).

As researchers uncover problematic access patterns that degrade the performance of the algorithms, many variants of LRU and LFU are devised to overcome the problems. 2Q~\cite{ccn101}, ARC~\cite{ccn102} and LIRS~\cite{ccn116} improve the performance by exploiting the advantages of LRU and LFU while their time and space complexities are comparable to that of LRU. In contrast to them, CLOCK~\cite{ccn100} reduces the complexity of LRU by approximating its behavior with a fixed circular buffer while keeping the performance. The complexity of CLOCK is comparable to that of NRU which has a low computational cost. CAR~\cite{ccn108} combines CLOCK with ARC to achieve both performance improvement and cost reduction.

Since web services became explosively popular, web-cache and CDN-cache are researched intensively to improve the performance of them in terms of bottleneck, latency, overload and robustness~\cite{ccn130,ccn131,ccn134}. Because the resource constraints of them are more moderate than that of computer systems, the cache replacement algorithms in a web and a CDN utilize statistical information including not only recency and frequency but also several others including size, latency, and URI~\cite{ccn131}. However, the improvement was slight or specific to particular environments in spite of an abundance of caching algorithms~\cite{ccn134}.

In recent years, ICN has revived research on caching algorithms because ICN provides inherent in-network caching feature. Unlike web- and CDN-cache employed in the application-layer, all elements in ICN have caching capability. Because one of the most interesting problems is improvement achieved by through cooperation among ICN elements in the network-layer, many researchers focus on cache placement algorithms~\cite{ccn127,ccn128}. As a cache replacement algorithm taking advantage of ICN, there are also policies that make use of content popularity~\cite{ccn105,ccn121}.

To realize ICN, especially an ICN core element, it is required to implement a cache replacement algorithm that can be operated with severe resource constraints instead of the statistical caching algorithms for web and CDN with rich resources. The implementation cost of commonly used approaches such as LRU and LFU are also prohibitive for router hardware, as pointed out by ~\cite{ccn120,ccn122}. Looking back at the history of cache replacement algorithms, ICN elements need a hardware implementable approach whose complexity is comparable to that of FIFO or CLOCK. In addition to the cost, this approach should cope with access patterns specific to ICN, where the unit of caching is a fine-grained chunk rather than whole content data. To understand how to satisfy these requirements of cost and performance, we examine the knowledge of caching in computer systems and apply it to in-network caching in the following section.

%%%%%%%%%%%%%%%%%%%%%%%%%%%%%%%%%%%%%%%%%%%%%%%%%%%%%%%%%%%%%%%%%%%%%%%%%%%%%%%%%%%%%%%%%%%%%%%%%%%%%%%%%%%%%%
\section{Design Considerations of Cache Replacement Algorithm for ICN}\label{sec:problem}

%%%%%%%%%%%%%%%%%%%%%%%%%%%%%%%%%%%%%%%%%%%%%%%%%%%%%%%%%%%%%%%%%%%%%%%%%%%%%%%%%%%%%%%%%%%%%%%%%%%%%%%%%%%%%%
\subsection{Access Patterns of Traffic in the Network}\label{sec:acs:ptn}

An access pattern is the important factor to govern the performance of cache replacement algorithm. It is well known that the popularity of contents follows a Zipf-like distribution: a large number of contents requested only once or just a few times~\cite{ccn300}. In addition, many requests generate asynchronous requests for contents, and so temporal locality of network traffic becomes relatively low.

In particular, ICN is able to identify a chunk (its default size is 4K bytes in CCNx), which enables the chunk level caching in an ICN element. Thus, we conjecture that the distribution of the ``chunk popularity" would be more biased than Zipf-like distributions. In this paper, to design an appropriate cache replacement algorithm for ICN under different types of the distributions, we classify access patterns of traffic, which governs the distribution~\cite{ccn102,ccn116,ccn101,ccn109}, into four categories: SCAN, LOOP, COOREALTED REFERENCES, and FICKLE INTEREST as follows: 
\begin{itemize}
    \item {SCAN: a sequence of requests to different chunks, and so each chunk is requested only once}
    \item {LOOP: a repetition of a scan}
    \item {CORRELATED REFERENCES: a short-term intensified requests to a few chunks}
    \item {FICKLE INTEREST: rapidly changing sets of requested chunks}
\end{itemize}

First, although the exact access pattern of the chunk level (i.e., network level) traffic in ICN is not known due to the lack of available ICN traffic trace, such one-time used items occupy 60\% or more in the network level traffic in IP networks~\cite{ccn060}. We conjecture that the highly aggregated network level traffic in ICN would have a large number of one-time used chunks, which correspond to SCAN access pattern.

Second, ICN is originally designed to efficiently disseminate multimedia traffic which generally occupies high network bandwidth and is requested repeatedly. Thus, we also conjecture that the chunk level traffic in ICN will have LOOP access pattern. As mentioned previously, LOOP is highly correlated to SCAN: SCAN and LOOP are generated by unpopular and popular contents, respectively. 

Third, CORRELATED REFERENCES and FICKLE INTEREST access patterns are observed in the requests to user-generated contents and real-time contents, respectively. We conjecture that these access patterns would be frequently observed in ICN due to the growth of social networks that share user-generated contents as well as real-time application such as video chatting. 
The volatile traffic hinders the strategies depending on statistical information (including LFU) from replacing the out-of-date chunks that were accessed frequently. 

For the reason above, the cache replacement algorithm for ICN should be able to deal with the access patterns described above. We here focus on the first access pattern, SCAN, in the design of the cache replacement algorithm for ICN since it is the major traffic that occupies the network bandwidth. Among the conventional cache replacement algorithms, CAR is able to efficiently deal with SCAN traffic access pattern~\cite{ccn108} due to its dual lists which enable to distinguish popular and non-popular contents. Our proposal is based on CAR to inherit this feature.

%%%%%%%%%%%%%%%%%%%%%%%%%%%%%%%%%%%%%%%%%%%%%%%%%%%%%%%%%%%%%%%%%%%%%%%%%%%%%%%%%%%%%%%%%%%%%%%%%%%%%%%%%%%%%%
\subsection{Computational Power and Memory Limitations}\label{sec:limit}

In the design of the cache replacement algorithm, two of the performance metrics should be considered. One is the cost that updates the table holding the information of cached items in the ICN element. The other is the cost that manages the table in the memory according to a cache replacement algorithm, e.g., prioritizing cached items. We call the former and the latter as computational cost and memory cost, respectively.

The computational cost includes insertion of a new caching item into the table, deletion of an existing cached item from the table, moving the location of cached items in the memory, and updating relevant information in the caching table. The operations listed above should be taken into account in the design of cache replacement algorithm, especially when it is applied for a high speed core ICN element.

The memory cost increases as the number of cached items increases due to the increase of control information for the maintenance of the table~\cite{ccn115,ccn108,ccn120,ccn122}. For example, LFU has much higher overhead to keep statistics of each cached item. In LRU using double-linked-list, this cost is prohibitive due to the maintenance of double pointers to other cached items.

To reduce the computational and memory costs in conventional approaches, CLOCK was introduced, which has a memory link list having a shape of a clock. It searches for a cached item that needs to be replaced following a clockwise. While searching for a candidate for replacement, it refers to one bit corresponding to the candidate. When the bit is set to off, the cached item is discarded. Otherwise, the searching process keeps on going until it finds a cached item whose bit is off. Then, all bits skipped over during the searching process are set to off. Thus, CLOCK requires only a single bit per chunk and few repetitions of the searching process. Our proposed mechanism also adopts this mechanism in CLOCK to reduce the computational and memory costs. 

%%%%%%%%%%%%%%%%%%%%%%%%%%%%%%%%%%%%%%%%%%%%%%%%%%%%%%%%%%%%%%%%%%%%%%%%%%%%%%%%%%%%%%%%%%%%%%%%%%%%%%%%%%%%%%
\subsection{Adaptable Parameter Tuning}\label{sec:param:tuning}

Some cache replacement algorithms need to tune parameters statically and dynamically according to workloads offered to the cached items in cache. For instance, the parameters include the interval to obtain statistics of request arrivals in LFU, the ratio between the number of popular and that of non-popular cached items in LIRS, and the variable sizes of the lists used in ARC and CAR.

While some parameters in ARC and CAR can be tuned adaptively to the change of workload, other parameters in LFU, LIRS and 2Q need to be defined in advance. However, the static parameters are unfavorable due to 1) difficulty of finding optimal parameters, 2) invalidity of optimal parameters in the change of workload, which causes performance fluctuation. For this reason, we conjecture that a cache replacement algorithm that adaptively changes the system parameters is preferable in the design of cache replacement algorithm for ICN.

%%%%%%%%%%%%%%%%%%%%%%%%%%%%%%%%%%%%%%%%%%%%%%%%%%%%%%%%%%%%%%%%%%%%%%%%%%%%%%%%%%%%%%%%%%%%%%%%%%%%%%%%%%%%%%
\section{Compact CLOCK with Adaptive Replacement (Compact CAR)}\label{sec:proposal}

%%%%%%%%%%%%%%%%%%%%%%%%%%%%%%%%%%%%%%%%%%%%%%%%%%%%%%%%%%%%%%%%%%%%%%%%%%%%%%%%%%%%%%%%%%%%%%%%%%%%%%%%%%%%%%
\subsection{Key Ideas of the Proposed Algorithm}\label{sec:key:idea}

The algorithm we propose is based on Clock with Adaptive Replacement (CAR)~\cite{ccn108}. CAR is known to be robust against the access patterns of SCAN, CORRELATED REFERENCES and FICKLE INTEREST.

CAR maintains two CLOCK lists: one is for cached items that have been accessed only once, and the other one is for the rest. CLOCK is a dominant algorithm in the design of page replacement in computer operating system~\cite{ccn108,ccn115}. Basically, it aims to achieve some of the benefits of LRU replacement but without heavy computational and memory costs in the manipulation of LRU operation - we discuss the operation in detail in the next section.

This two-stack approach can keep frequently requested contents from being replaced by one-time requested content. In addition, CAR can dynamically adapt to the access patterns by adjusting its lengths of CLOCK lists as autonomously tuning parameter. Due to the adaptive parameter tuning, CAR can also function well in an unsteady environment. However, the computational cost of CAR is similar to that of LRU, which is prohibitive in the use of the algorithm for the design of high speed ICN core element.

%%%%%%%%%%%%%%%%%%%%%%%%%%%%%%%%%%%%%%%%%%%%%%%%%%%%%%%%%%%%%%%%%%%%%%%%%%%%%%%%%%%%%%%%%%%%%%%%%%%%%%%%%%%%%%
\subsection{Design of Compact CAR}\label{sec:design}

\begin{figure*}[t!]
\begin{center}
\includegraphics[width = 0.70\hsize,clip]{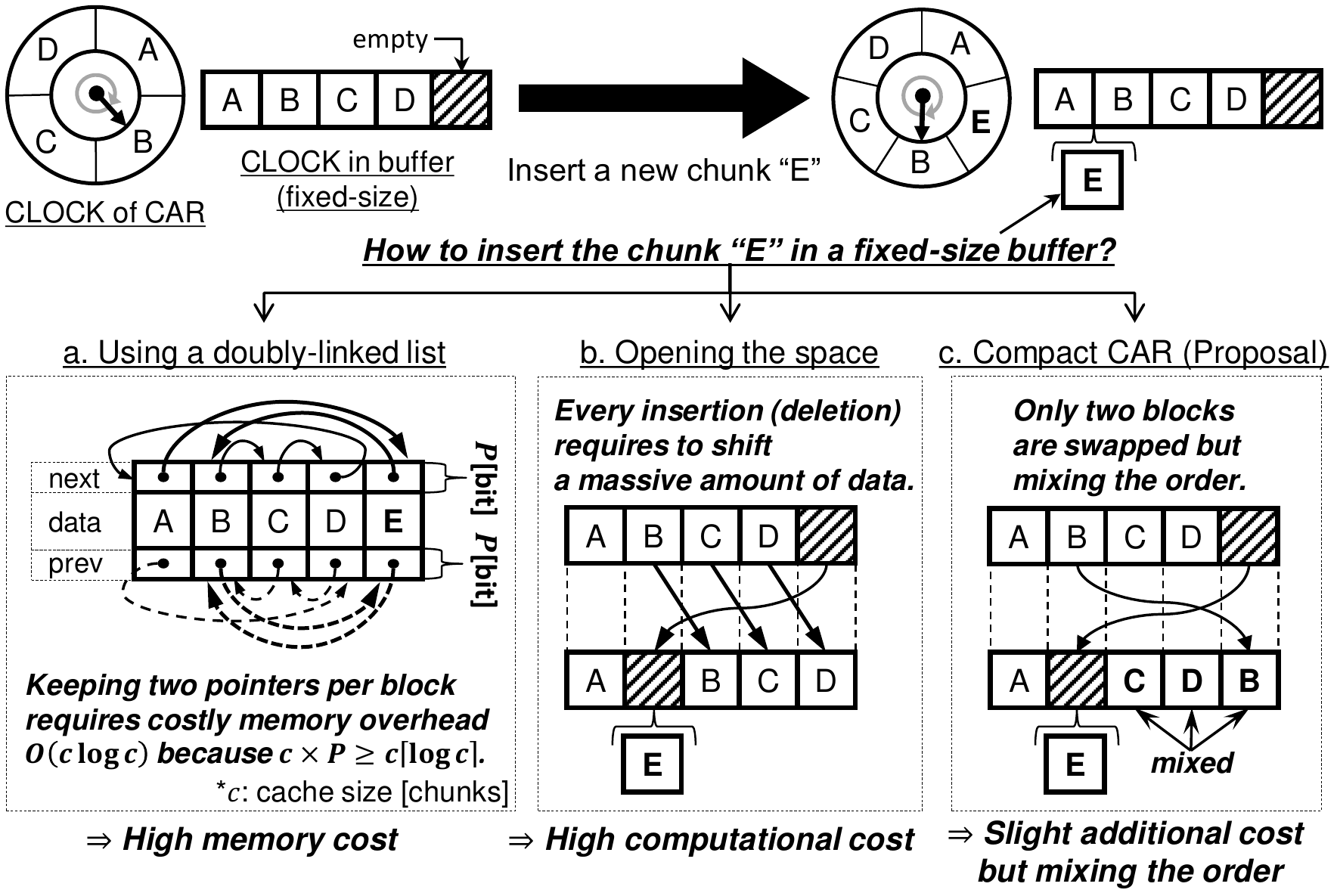}
\caption{Illustration of Computational and Memory Costs in the Inserting Operation in the Different Data Structures}
\label{cost:insertion:car}
\end{center}
\end{figure*}

\begin{figure*}[t!]
\begin{center}
\includegraphics[width = 0.70\hsize,clip]{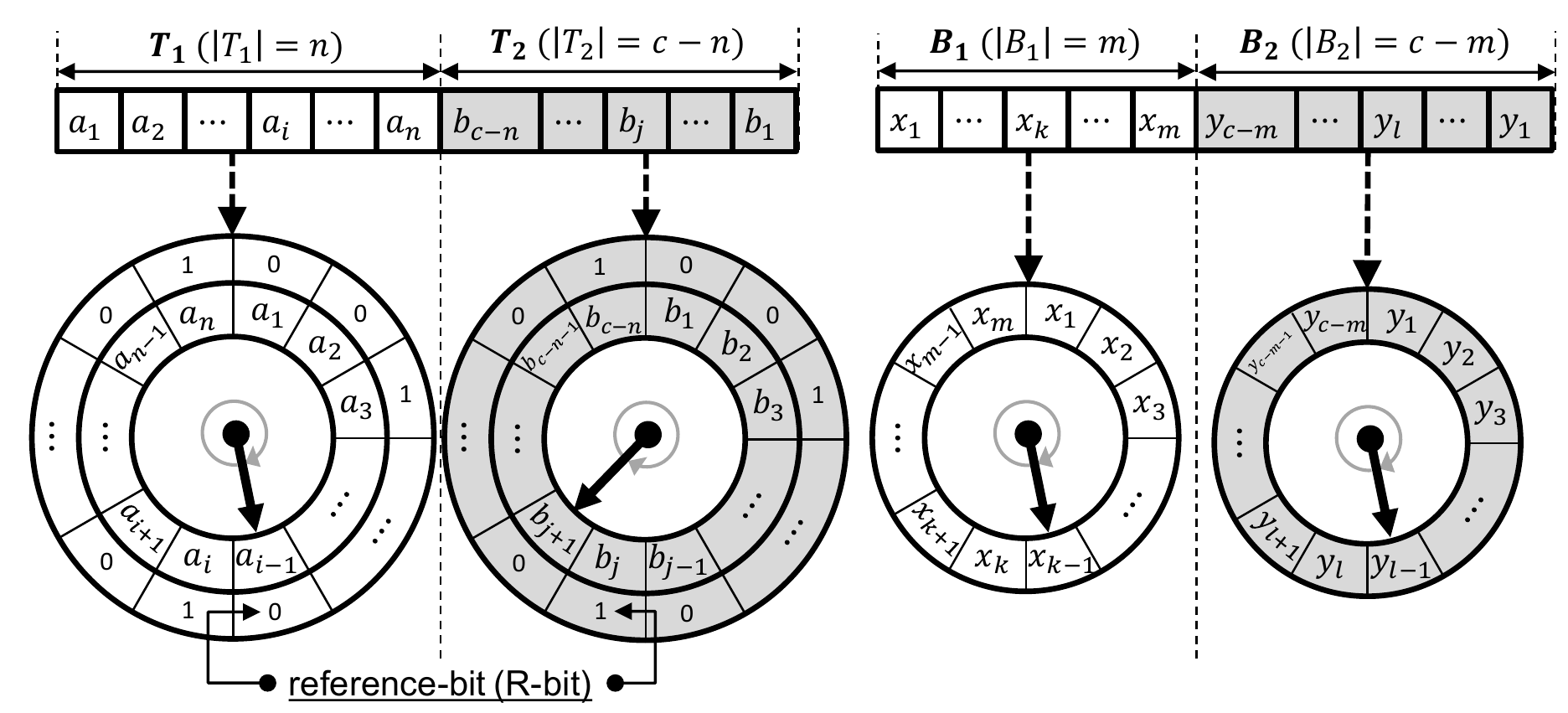}
\caption{Data Structure of Compact CAR}
\label{arclock:shift:border:a}
\end{center}
\end{figure*}

\begin{figure*}[t!]
\begin{center}
\includegraphics[width = 0.75\hsize,clip]{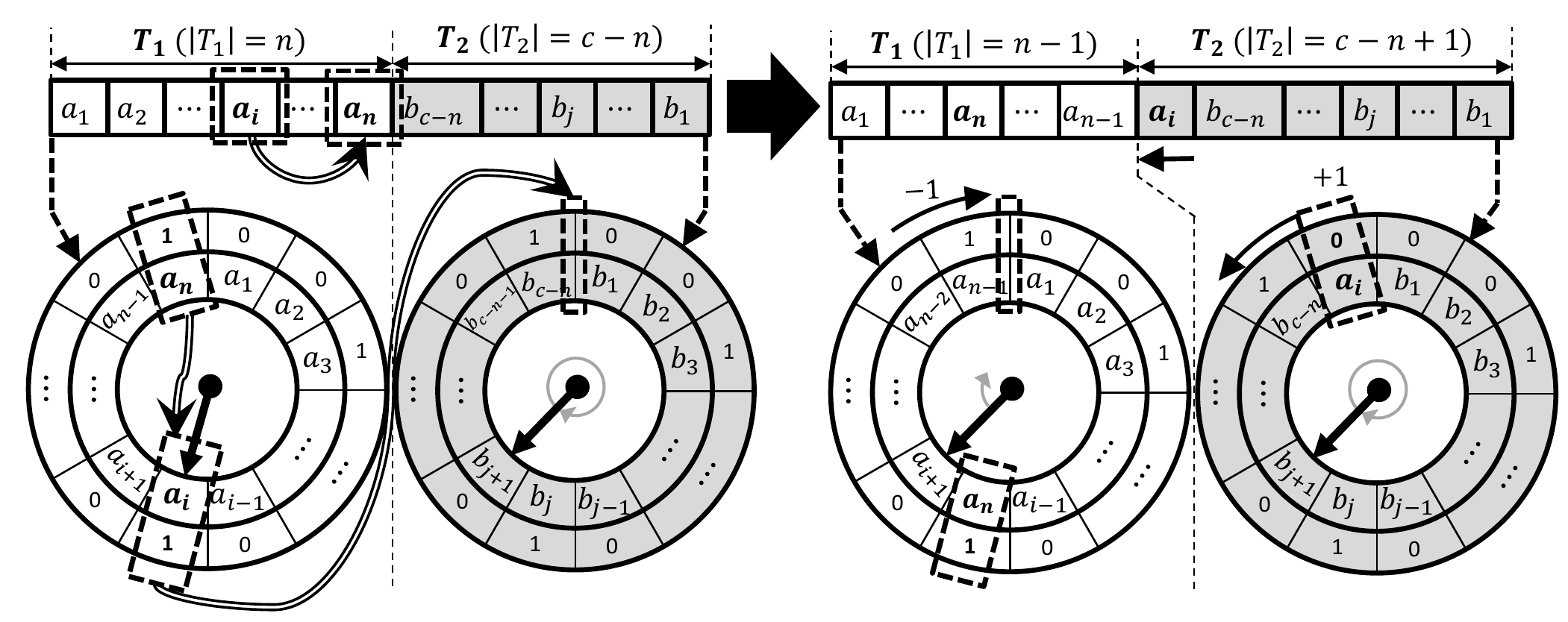}
\caption{Example of Moving a Chunk $a_i$ from $T_1$ to $T_2$ by Replacing the Edge Chunk $a_n$}
\label{arclock:shift:border:b}
\end{center}
\end{figure*}

Compact CAR is designed to further reduce the memory cost of CAR while maintaining its inherent advantages. The reason that CAR has high memory cost is that the sizes of CLOCK lists used in CAR have variable-size. Thus, Compact CAR has the two CLOCK lists in the fixed-size buffer to overcome the problem.

Figure \ref{cost:insertion:car} illustrates an operation of chunk insertion with different approaches from the viewpoint of computational and memory costs. First, CAR uses a doubly-linked list. When a chunk is inserted in the middle of memory space, the chunk is inserted physically at the end of the memory space. Then, the order of the chunks in the memory is arranged virtually using a doubly-linked list. It involves with two operational costs: computational cost which involves the rearrangement of pointers in the doubly-linked list, and memory cost which involves the memory space accommodating the doubly-linked list. Computational cost is not that expensive. However, it consumes a decent amount of memory space to maintain the order by keeping two pointers per block.

Second, it illustrates a case where a doubly-linked list is not used but memory blocks are shifted when a chunk is inserted. It does not require high memory cost because the order of chunks in physical memory can be used directly without creating virtual order created in the first scenario. However, this scenario demonstrates high computational cost caused by the shift of memory blocks.

Third, the operation of chunk insertion in Compact CAR is illustrated. To insert a new chunk at the position of the chunk 'B', Compact CAR moves the chunk 'B' to the end. Then, the newly inserting chunk is inserted to the location. It does not use a doubly-linked list to create virtual order of chunks in the memory space and so the memory cost can be reduced dramatically. At the same time, it does not involve the shift of memory blocks simultaneously, which reduces a computational cost as well. Readers may concern that the mixed order degrades the performance but it is not that serious: we will address the issue in Section \ref{sec:eval}.

Figure \ref{arclock:shift:border:a} illustrates the data structure of Compact CAR which maintains four variable-size CLOCK lists. Let us say $T_*=T_1\cup T_2$ and $B_*=B_1\cup B_2$. $T_*$ are used for caching data, and $B_*$ remember the record of evicted chunks. Each of $T_*$ and $B_*$ is arranged in physically contiguous memory. The $n$ chunks in $T_1$ are arranged in one side and the $(c-n)$ chunks in $T_2$ fill the other side. To realize the searching process described in Section \ref{sec:limit}, $T_*$ and $B_*$ have a hand as the starting position to start for searching process, and $T_*$ assigns each chunk a reference bit ($R$-bit). The hand of $T_1$ moves rightward on the array in the figure and that of $T_2$ moves leftward so as to fill the free space in the direction that the hand moves (i.e., if cache is not full, there are empty blocks between $T_1$ and $T_2$). The same principle applies to $B_1$ and $B_2$, except that they do not hold data of chunks and $R$-bits. Compared to the operation of CAR in the related work, Compact CAR does not require a memory space to keep pointers as discussed above. This is a reason why Compact CAR can use memory more efficiently in a compact manner than CAR.

As mentioned in Section \ref{sec:key:idea}, Compact CAR inherits advantages of two-stack approach of CAR. $L_1=T_1\cup B_1$ (unshaded) and $L_2=T_2\cup B_2$ (shaded) are assumed to be two CLOCK lists. $L_1$ is for contents that have been accessed only once. $L_2$ is for contents that have been accessed at least twice. By adjusting the target size $p$, which is the parameter representing the target size of $T_1$ ($0< p\leq c$), the sizes of the lists $|T_1|, |B_1|, |T_2|$, and $|B_2|$ vary adaptively. A cache hit to $L_1$ ($L_2$) dynamically increases the target size for $T_1$ ($T_2$), and simultaneously decreases the target size for the other list. In other words, the size of $T_1$ grows when the recency of contents governs the performance of the cache replacement algorithm, whose behavior becomes similar to $LRU$. On the other hand, the size of $T_2$ grows when the frequency of content access governs the performance of the cache replacement algorithm, whose behavior becomes similar to $LFU$. For example, the size of $T_2$ grows when SCAN occurs. This feature enables Compact CAR to adaptively deal with the change of traffic access patterns.

Figure \ref{arclock:shift:border:b} gives an example of moving a chunk $a_i$ within $T_*$ to realize the swap operation illustrated in Fig. \ref{cost:insertion:car}(c). Compact CAR swaps $a_i$ with the one at the boundary between $T_1$ and $T_2$, and then just shifts the boundary to the left hand side. This simple swap operation enables the migration of a chunk between $T_1$ and $T_2$. However, the original CAR algorithm requires to keep a doubly linked list or to shift an enormous amount of chunks $(a_{i+1}, \cdots , a_n, b_{c-n}, \cdots, b_{j+1})$ leftward to insert $a_i$ into the position pointed by the hand of $T_2$ as illustrated in Figure \ref{cost:insertion:car}.

\subsection{Replacement Algorithm of Compact CAR}\label{sec:algorithm}

Algorithms 1, 2, 3, and 4 show pseudocode of the cache replacement algorithm of Compact CAR. The replacement process in cache replacement starts with Algorithm 1. If an accessed chunk $x$ is hit, then the process sets the $R$-bit of $x$ and terminates (lines 2--4). If there is a cache miss, then line 5 checks whether $x$ is in a ghost cache. If $B_*$ contains $x$, the history of $x$ is discarded and the parameter $p$, which is the target size for $T_1$, is updated (lines 7--13). This process of tuning $p$ makes CAR adaptive to changes in access patterns. If $x$ is not in $B_*$ and the ghost cache if full, then a chunk in $B_*$ is discarded (lines 16--18). In this, $i$ stands for the index of the list into which $x$ is to be cached; therefore, $i$ is set to 2 when there is a ghost hit and is set to 1 otherwise. After ensuring that there is room in the ghost cache, a chunk in $T_*$ is discarded if $T_*$ is full (lines 20--23). A victim chunk is selected from $T_1$ if the size of $T_1$ is not less than the target size $p$; otherwise, a chunk in $T_2$ is replaced. Finally, $x$ is cached at a position $s_t$ in $T_i$, which has been ensured to be free (line 27).

Algorithms 2, 3, and 4 describe how to make room for a chunk or its historical information. ${\rm Hand}_{T_i}$ indicates the location pointed to by the hand of $T_i$, and $B_i$ is analogous. Every algorithm assures a free space at {\it EdgeAddr}, which is the address of the boundary of two lists, either $T_1$ and $T_2$ or $B_1$ and $B_2$. Whenever $x$ is not located at the boundary, it is swapped with the {\it EdgeChunk} of the list, which is located at the boundary. The reason to do so is that this frees an address located next to the edge address of both lists. For example, consider what happens when caching a new chunk in $T_1$ when a chunk in $T_2$ is discarded. If the cache is full and the swap is not performed, there is no free space contiguous with the area of $T_1$ unless the discarded chunk in $T_2$ is adjacent to $T_1$. Otherwise, it will be necessary to swap two chunks or shift an enormous amount of chunks in order to cache the new chunk in $T_1$. Thus, the swap process is essential to ensure a vacant address that is contiguous with both of the lists (not only $T_1$ and $T_2$, but also $B_1$ and $B_2$). 

The \texttt{DiscardBtm} procedure (shown in Algorithm 2) evicts the historical information of $x$ in $B_i$ when $x$ is a ghost hit. The \texttt{ReplaceBtm} procedure (shown in Algorithm 3) evicts the history information pointed to by the hand of $B_i$ when there is a cache miss and $B_i$ is full. The \texttt{ReplaceTop} procedure (shown in Algorithm 4) removes the chunk pointed by the hand of $T_i$ when the cache is full. If a chunk whose $R$-bit is set is found in $T_1$, the chunk is moved to $T_2$ in the manner described in Fig.5. Lines 11--12 store the evicted chunk as history information without losing the contiguousness of the lists because an address next to the edge of $B_i$ has been ensured to be free, as discussed above.

{%\begin{small}
\small
\renewcommand{\baselinestretch}{0.75}
\begin{algorithm}[t!]
\caption{Compact CAR Replacement Algorithm}
\label{alg:car:rep}
\begin{algorithmic}[1]
\Procedure{CacheReplacement}{$x$} \Comment{$x$ is an accessed chunk.}
	\If{$x \in T_{*}$} \Comment{cache hit}
		\State $x.$R-bit$\gets 1$
		\State {\bf return}
	\ElsIf{$x \in B_{*}$} \Comment{ghost hit}
		\State $i\gets 2$ \Comment{to cache $x$ in $T_2$}
		\If{$x \in B_1$}
			\State $\delta \gets {\rm max}(1, \frac{|B_2|}{|B_1|})$; $p\gets {\rm min}(c, p + \delta)$
			\State DiscardBtm(1,$x$)
		\Else \Comment{$x \in B_2$}
			\State $\delta \gets {\rm max}(1, \frac{|B_1|}{|B_2|})$; $p\gets {\rm max}(0, p - \delta)$
			\State DiscardBtm(2,$x$)
		\EndIf
	\Else \Comment{cache miss}
		\State $i\gets 1$ \Comment{to cache $x$ in $T_1$}
		\If{${\rm Full}(L_1) \And |B_1|>0$} ~ ${\rm ReplaceBtm}(1)$
		\ElsIf{${\rm Full}(L) \And |B_2|>0$} ${\rm ReplaceBtm}(2)$
		\EndIf
	\EndIf
	\If{${\rm Full}(T_*)$}
		\If{$|T_1| \geq {\rm max}(p,1)$} ~ $s_t\gets {\rm ReplaceTop}(1)$
		\Else ~ $s_t\gets {\rm ReplaceTop}(2)$
		\EndIf
	\Else \Comment{$T_*$ is not full.}
		\State $s_t\gets $ an available address in $T_i$
	\EndIf
	\State $T_i[s_t]\gets x$ \Comment{$x$ is cached as a chunk in $T_i$.}
\EndProcedure
\end{algorithmic}
\end{algorithm}
}%end{small}

{%\begin{small}
\small
\renewcommand{\baselinestretch}{0.75}
\begin{algorithm}[t!]
\caption{DiscardBtm() for Compact CAR}
\label{alg:car:compact:discard:b}
\begin{algorithmic}[1]
\Procedure{DiscardBtm}{$i,x$}
	\State Swap$(x,B_i.{\rm EdgeChunk})$
	\State Discard $x$ (at the edge of $B_i$)
	\State \Comment{ensuring that an address next to the edge of $B_i$ is free}
\EndProcedure
\end{algorithmic}
\end{algorithm}
}%end{small}

{%\begin{small}
\small
\renewcommand{\baselinestretch}{0.75}
\begin{algorithm}[t!]
\caption{ReplaceBtm() for Compact CAR}
\label{alg:car:compact:replace:b}
\begin{algorithmic}[1]
\Procedure{ReplaceBtm}{$i$}
	\State Swap$(B_i[{\rm Hand}_{B_i}],B_i.{\rm EdgeChunk})$
	\State Discard $B_i.{\rm EdgeChunk}$
	\State Rotate ${\rm Hand}_{B_i}$
	\State \Comment{ensuring that an address next to the edge of $B_i$ is free}
\EndProcedure
\end{algorithmic}
\end{algorithm}
}%end{small}

{%\begin{small}
\small
\renewcommand{\baselinestretch}{0.75}
\begin{algorithm}[t!]
\caption{ReplaceTop() for Compact CAR}
\label{alg:car:compact:replace:t}
\begin{algorithmic}[1]
\Function{ReplaceTop}{$i$}
	\While{$T_i[{\rm Hand}_{T_i}].$R-bit$ = 1$}
		\State $T_i[{\rm Hand}_{T_i}].$R-bit$ \gets 0$
		\If{i=1}
			\State Swap$(T_i[{\rm Hand}_{T_i}],T_i.{\rm EdgeChunk})$
			\State \Comment{Shift the boundary between $T_1$ and $T_2$.}
		\EndIf
		\State Rotate ${\rm Hand}_{T_i}$
	\EndWhile
	\State $s_{e} \gets$ an address next to the edge of $B_i$
	\State $B_i[s_{e}]\gets T_i[{\rm Hand}_{T_i}]$
	\State Swap$(T_i[{\rm Hand}_{T_i}],T_i.{\rm EdgeChunk})$
	\State Discard $T_i.{\rm EdgeChunk}$
	\State Rotate ${\rm Hand}_{T_i}$
	\State {\bf return} $T_i.{\rm EdgeAddr}$
\EndFunction
\end{algorithmic}
\end{algorithm}
}%end{small}

%%%%%%%%%%%%%%%%%%%%%%%%%%%%%%%%%%%%%%%%%%%%%%%%%%%%%%%%%%%%%%%%%%%%%%%%%%%%%%%%%%%%%%%%%%%%%%%%%%%%%%%%%%%%%%
\section{Performance Evaluation}\label{sec:eval}

In this section, we evaluate the performance of Compact CAR compared to OPT (off-line optimal algorithm with a priori knowledge of the stream of requests: absolute upper bound on the achievable cache hit rate), FIFO, CLOCK, and CAR in various scenarios to demonstrate the fulfillment of the design considerations discussed previously. 

First, the performance of the proposed algorithm is evaluated with various access patterns including synthetic traffic as well as real traffic trace in different types of topologies. Then, adaptability of our proposal to changing access traffic patterns is demonstrated by comparing to the same approach without tuning a parameter. Finally, computational and memory costs of the proposal are theoretically analyzed to present its efficient memory usage which is critical in the design of a high performance ICN core element.

%%%%%%%%%%%%%%%%%%%%%%%%%%%%%%%%%%%%%%%%%%%%%%%%%%%%%%%%%%%%%%%%%%%%%%%%%%%%%%%%%%%%%%%%%%%%%%%%%%%%%%%%%%%%%%
\subsection{Simulation Setup and Configuration}\label{sec:setting}

Two types of workloads are used in this simulation study: artificial workloads that follow a Zipf distribution and real traffic traces of Video-on-Demand (VoD), e.g., YouTube, DailyMotion, and NicoVideo, which are collected from a network gateway at Osaka University campus. The former and the latter are denoted by $A_{Zipf(\alpha)}$ and by $A_{Real}$. In addition, their superscript $C$ and $P$, e.g., $A_{Zipf(a)}^{C}$ and $A_{Zipf(\alpha)}^{P}$ represent the sizes of content and chunk, respectively.

The popularity of Internet content (e.g., VoD, web pages, file sharing, and user generated traffic) has been reported to follow the Zipf distribution with $0.6\leq \alpha \leq 1.2$~\cite{ccn019,ccn060}. Thus, we use these values to generate synthetic traffic requests from the Zipf distribution for this simulation study.

\begin{figure}[t!]
\begin{center}
\includegraphics[width=\columnwidth ,clip]{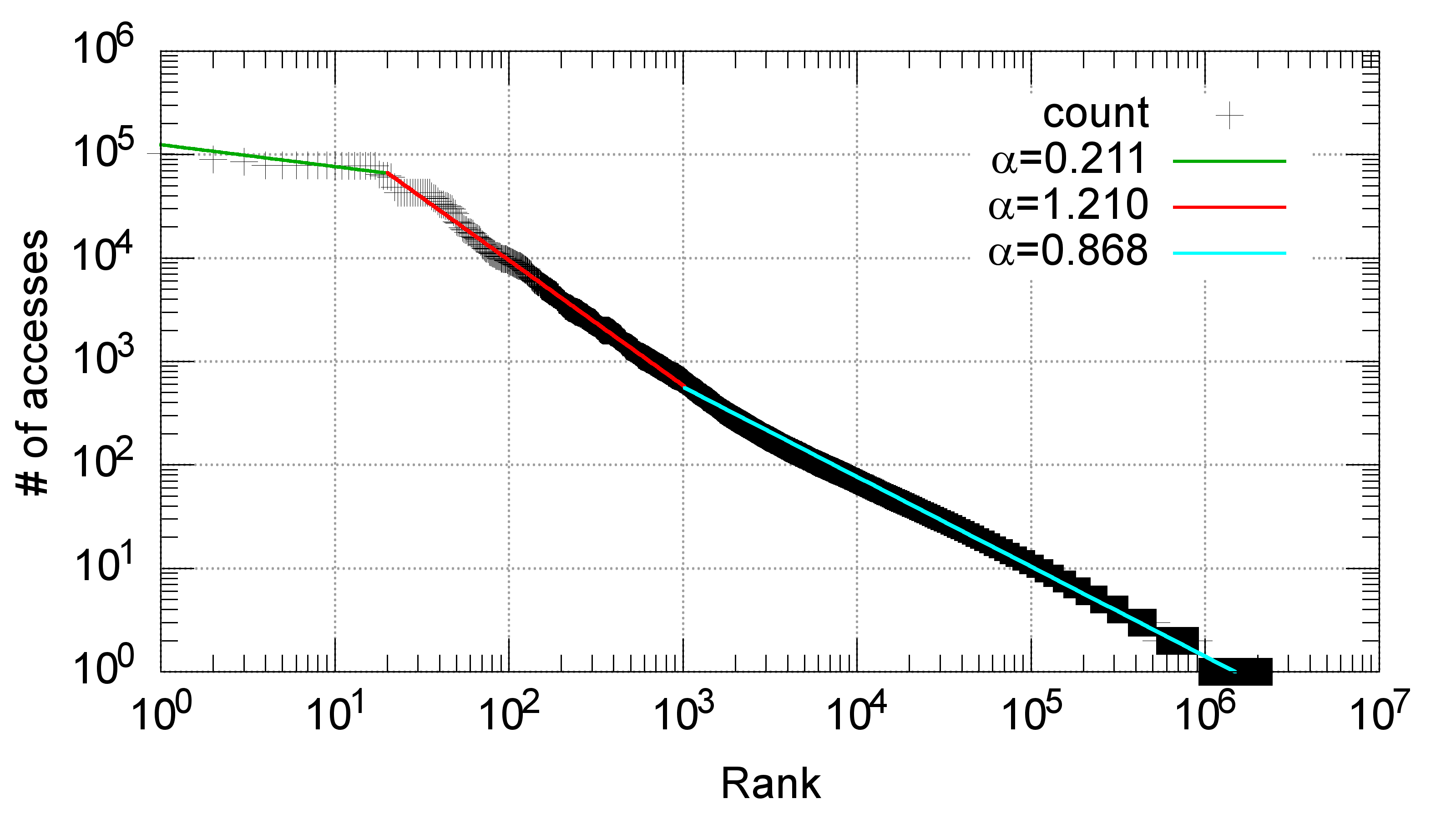}
\caption{Popularity Distribution of Real Trace}
\label{actual:access:dist:zipf}
\end{center}
\end{figure}

At the same time, the real traffic traces are gathered from July 26th 2013 to February 26th 2015. The number of unique contents is 2,428,880; the number of contents requested at least twice is 918,545; and the number of total accesses is 13,004,868. The popularity distribution of the real traffic trace follows the Zipf-like distribution, as depicted in Fig. \ref{actual:access:dist:zipf}. We also show the statistics of the real traffic traces in units of chunks in Table \ref{stat:of:workload:chunk}.

As stated in Section \ref{sec:acs:ptn}, the fine granularity of cacheable items in ICN, namely chunks or segments, changes the access patterns of request message, which dramatically governs the performance of cache replacement algorithm. Unfortunately, ICN traffic traces are not available yet. Thus, we generate synthetic requests for chunks, which simulates the access pattern of ICN in the following manner. We assume that the inter arrival time between requests to content items is similar to the one in the current Internet, which follows the Zipf distribution. However, the inter arrive time for chunks is constant according to Table \ref{rule:of:packet:stream:yt}, which is determined by the statistics of our observed real traffic. The generated requests are superimposed to simulate the aggregation of request messages in the network.

Two different topologies are used for this simulation study. One is a topology in which there is only one ICN element between clients and a server. The other is a line topology that includes ten ICN elements between them. 
One ICN element topology is used to demonstrate the performance of the proposed algorithm compared to those of conventional approaches including an optimal performance.
On the other hand, the line topology is used to present how efficiently the proposed cache replacement algorithm works without a cooperative caching mechanism\footnote{It cooperatively distributes content items in the network to improve cache hits as well as to reduce the usage of network resources.}.
In ICN, a requested content can be cached while being downloaded at any node along the downloading path, which is known as on-path caching. 
When all nodes along the path cache downloading contents, it eliminates the effect of a cooperative caching.
In other words, the simulation results with a line topology reveal how resistant our proposed algorithm in the absence of cooperative caching mechanism.

Each cache at ICN element has same capacity $c$ of $10^1$ to $10^6$ chunks which are adjusted according to the traffic trace we adopt. Also, the transmission delay of each chunk on links and the unnecessary computation in the protocol stacks are ignored to simplify the simulation.

\begin{table}[t!]
\footnotesize
\begin{center}
\caption{Number of Chunks Per Second $[{\rm pck}/{\rm s}]$}
\label{rule:of:packet:stream:yt}
\begin{tabular}{|c|c|c|c|}
\hline
Chunk size   & 1.5 KB & 15 KB & 60 KB \\
\hline
Standard Definition ($600 {\rm kbps}$) &  50 &  5 & 1.25 \\
High Definition ($1.2 {\rm Mbps}$)     & 100 & 10 & 2.50 \\
\hline
\end{tabular}
\end{center}
\end{table}

\begin{table}[t!]
\footnotesize
\begin{center}
\renewcommand{\arraystretch}{1.5}
\caption{Statistics of Workloads in Units of Chunks}
\label{stat:of:workload:chunk}
\begin{tabular}{|c|r|r|r|}
\hline
Workload & \multicolumn{1}{|p{13mm}|}{\# of total accesses} & 
           \multicolumn{1}{|p{18mm}|}{\# of observed unique chunks} & 
           \multicolumn{1}{|p{15mm}|}{\# of chunks requested at least twice} \\
\hline
\acsYPa  & 17,955,409 & 5,465,044 &   440,254 \\
\acsYPb  & 14,557,548 & 5,321,617 &   552,631 \\
\acsYPc  & 16,606,810 & 8,006,084 & 1,769,759 \\
\hline
\end{tabular}
\end{center}
\end{table}

%%%%%%%%%%%%%%%%%%%%%%%%%%%%%%%%%%%%%%%%%%%%%%%%%%%%%%%%%%%%%%%%%%%%%%%%%%%%%%%%%%%%%%%%%%%%%%%%%%%%%%%%%%%%%%
\subsection{Cache Hit Rate with Synthetic Traffic}\label{sec:eval:zipf}

\begin{figure*}[t!]
\subgraphcachehitrate{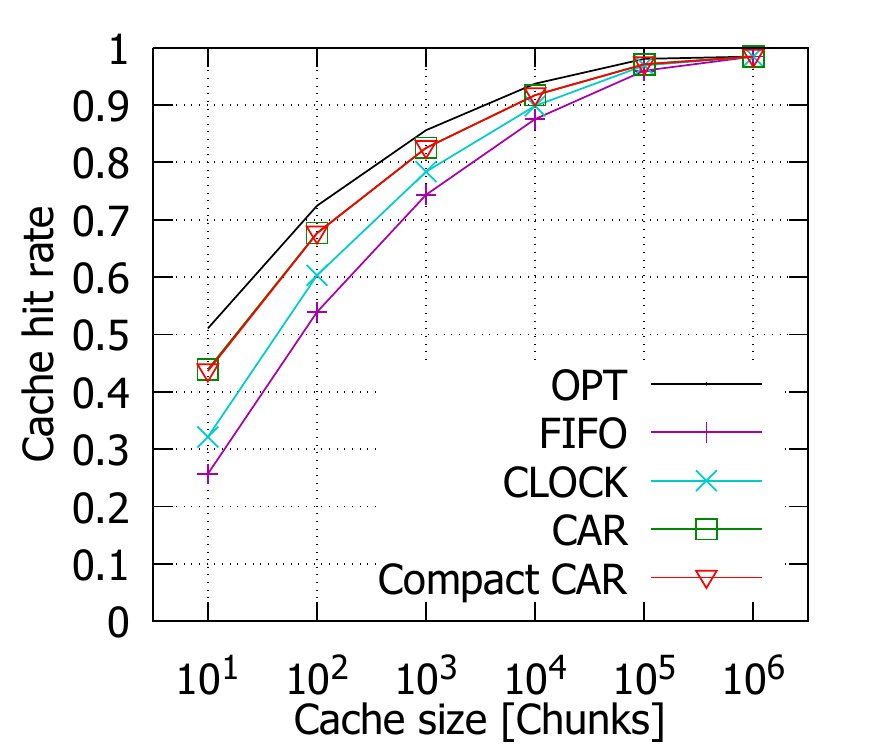}{\acsZdC}{car:acszdc}
\subgraphcachehitrate{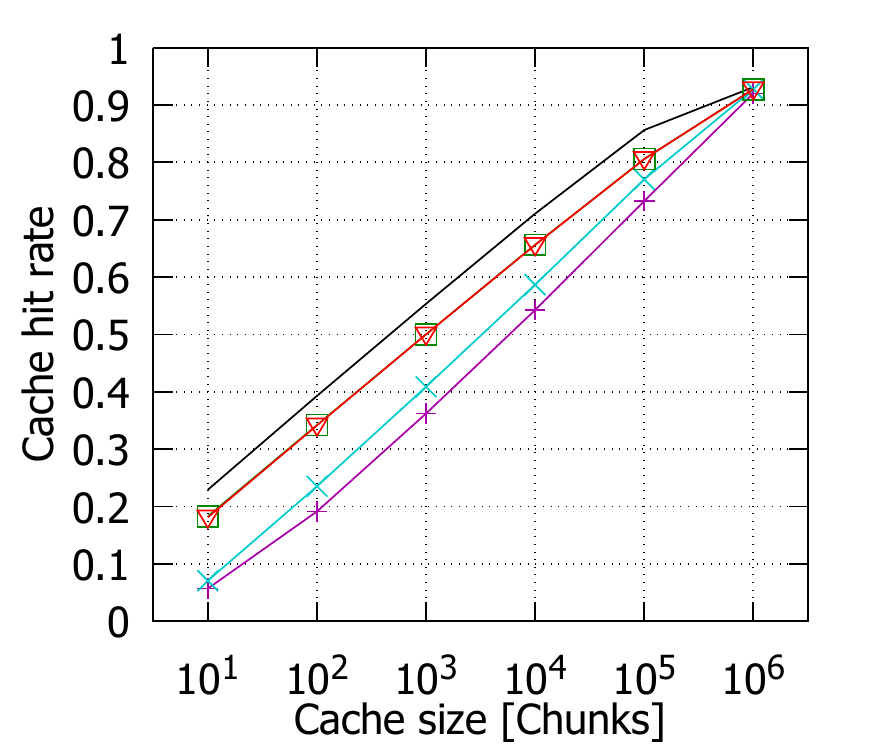}{\acsZcC}{car:acszcc}
\subgraphcachehitrate{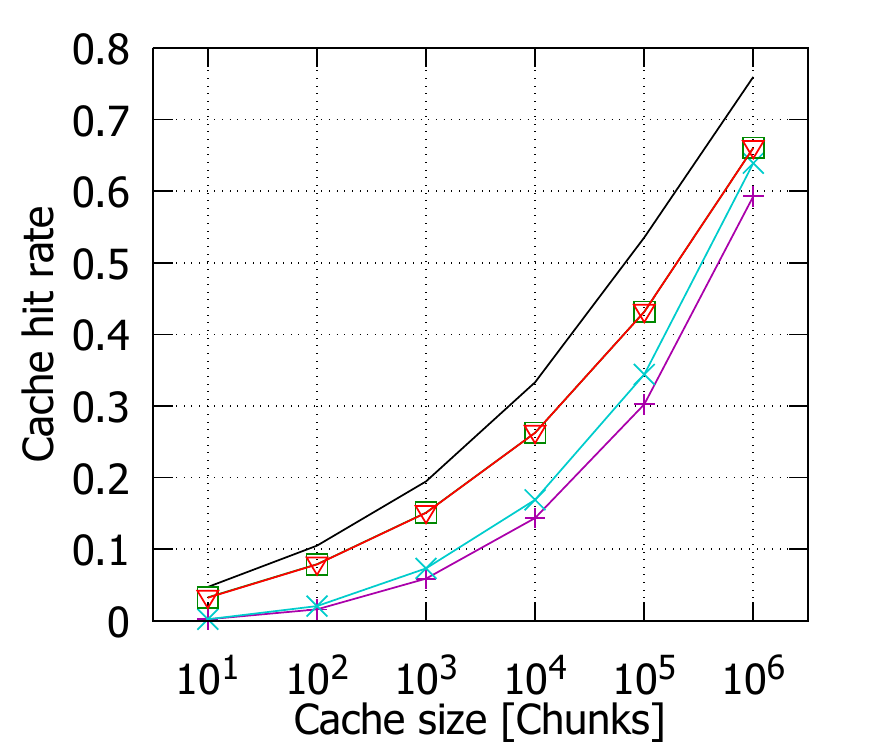}{\acsZbC}{car:acszbc}
\subgraphcachehitrate{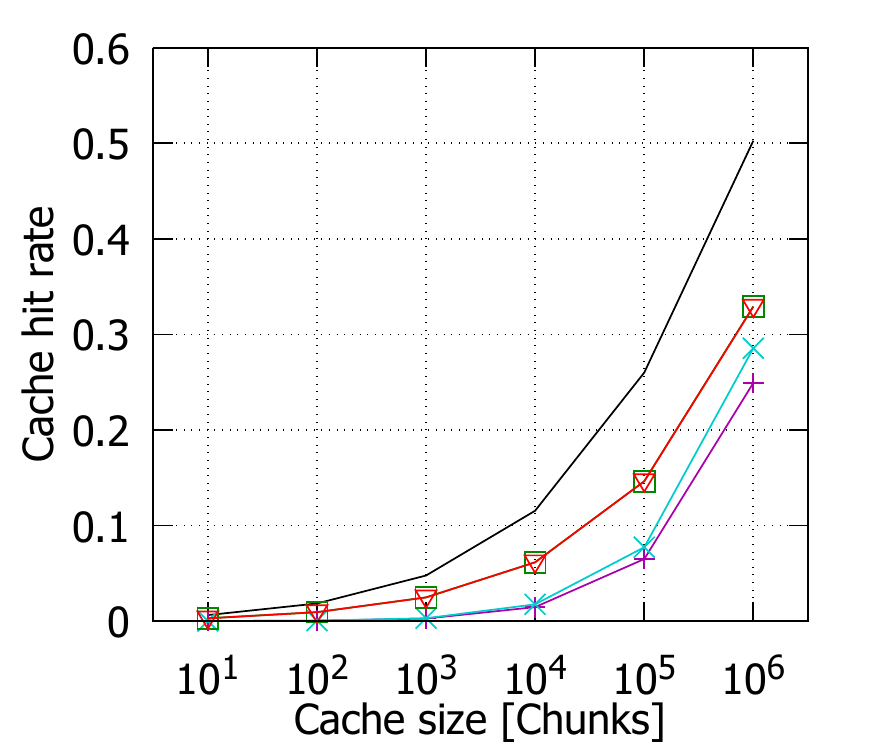}{\acsZaC}{car:acszac}
\caption{Results for Artificial Workloads in Units of Content}
\label{cache:hit:rate:car:acszc}
\end{figure*}

\begin{figure*}[t!]
\subgraphcachehitrate{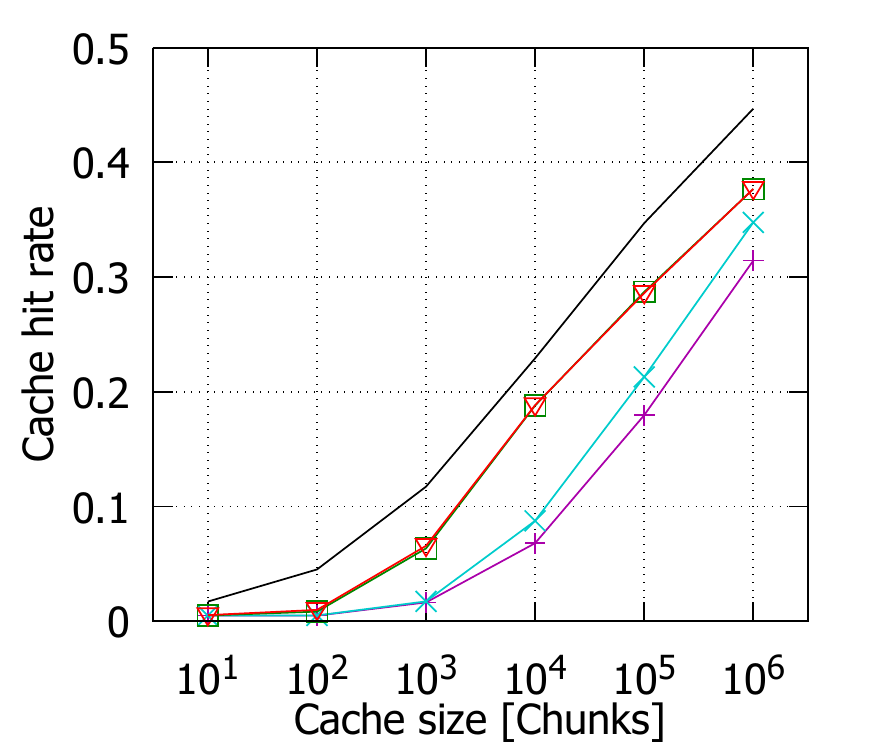}{\acsZcPc}{car:acszcpc}
\subgraphcachehitrate{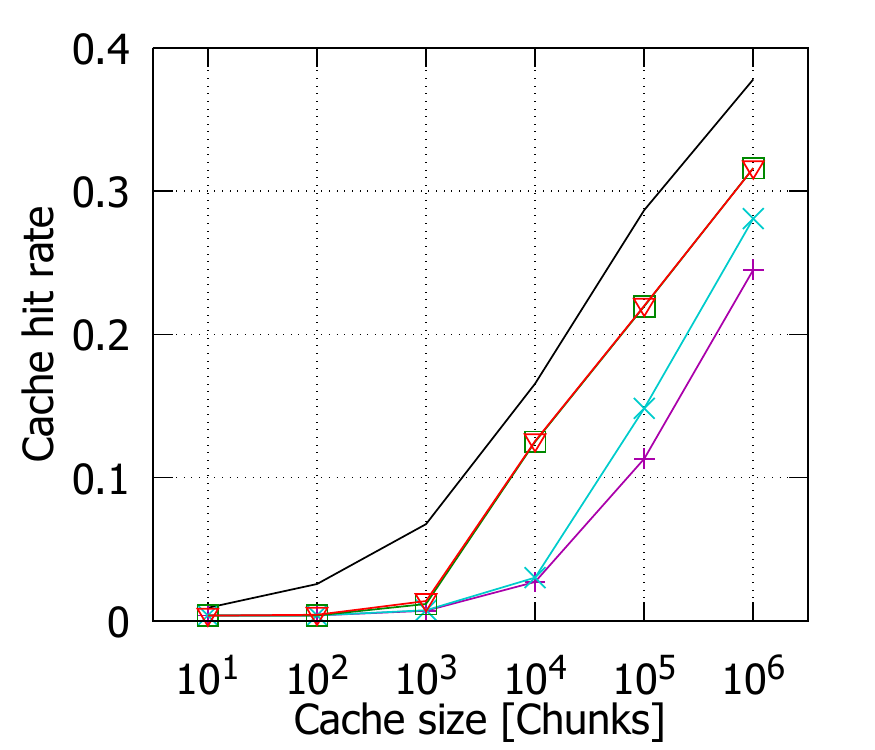}{\acsZcPb}{car:acszcpb}
%\subgraphcachehitrate{CAR_single_z_100_p_1500}{\acsZcPa}{car:acszcpa}
\subgraphcachehitrate{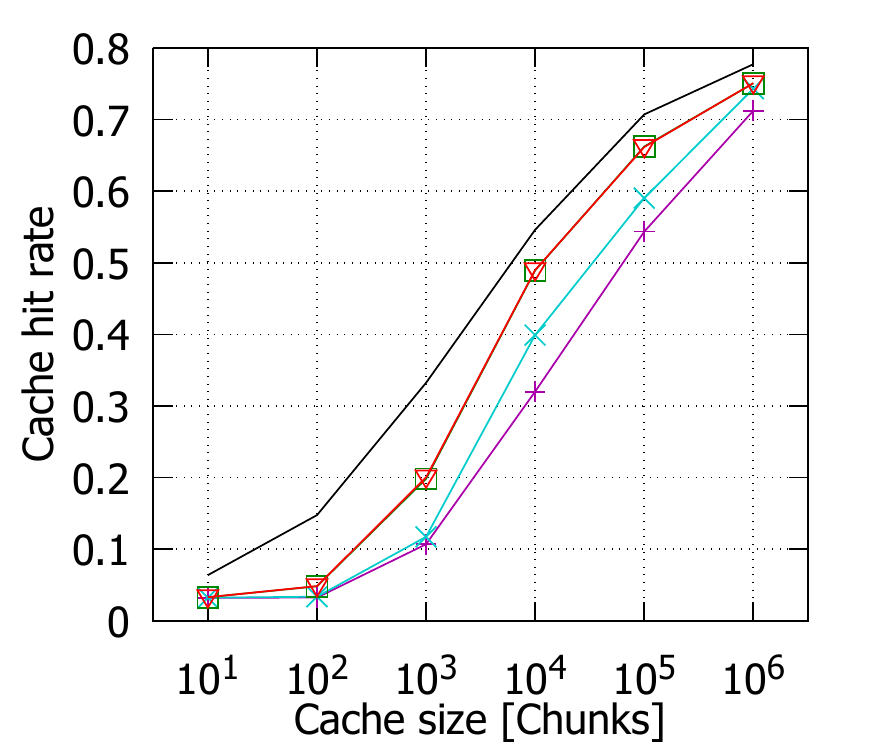}{\acsZdPc}{car:acszdpc}
\subfigure[$A_{{\rm Zipf}(1.0)}$, $c=10^5$]{\includegraphics[clip, width=0.5\columnwidth ]%
{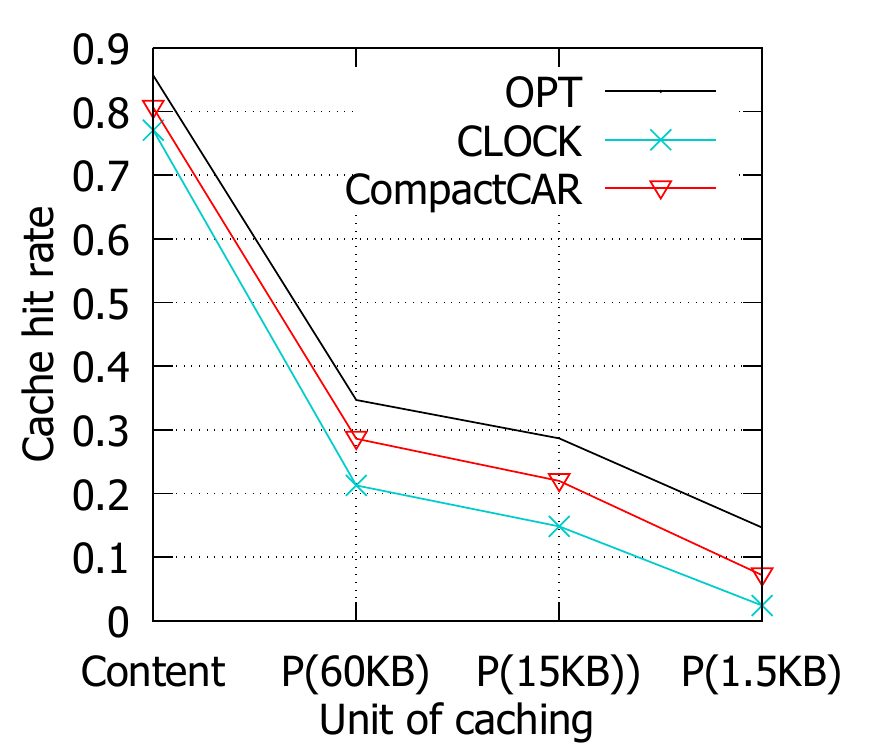}%
\label{cache:hit:rate:car:acszc:ctop}}
\caption{Results for Artificial Workloads in Units of Chunks}
\label{cache:hit:rate:car:acszp}
\end{figure*}

\begin{figure*}[t!]
\subgraphcachehitrate{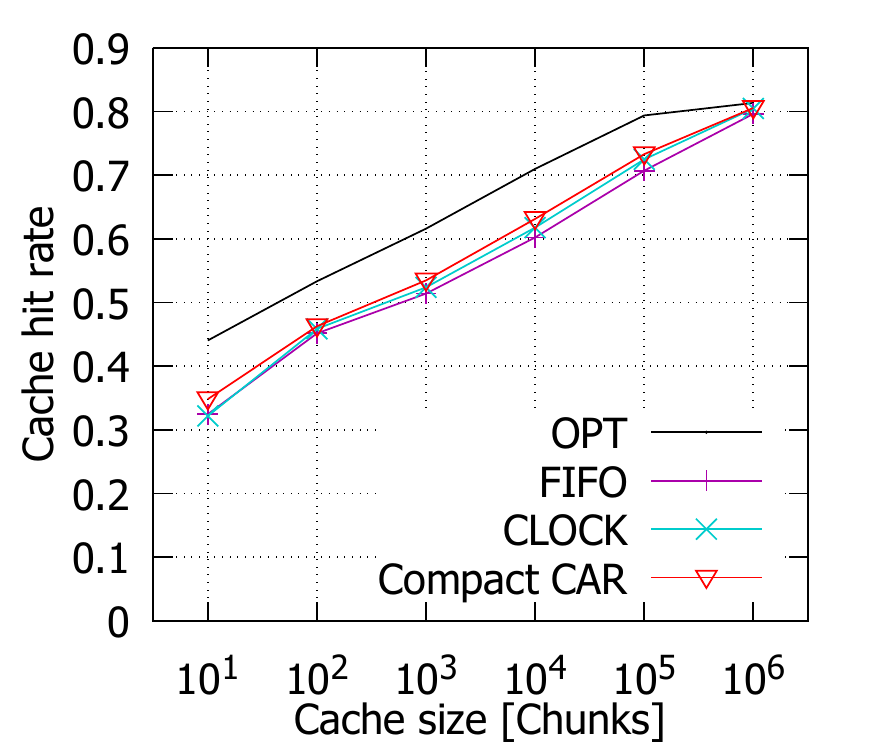}{\acsYC}{car:acsyc}
\subgraphcachehitrate{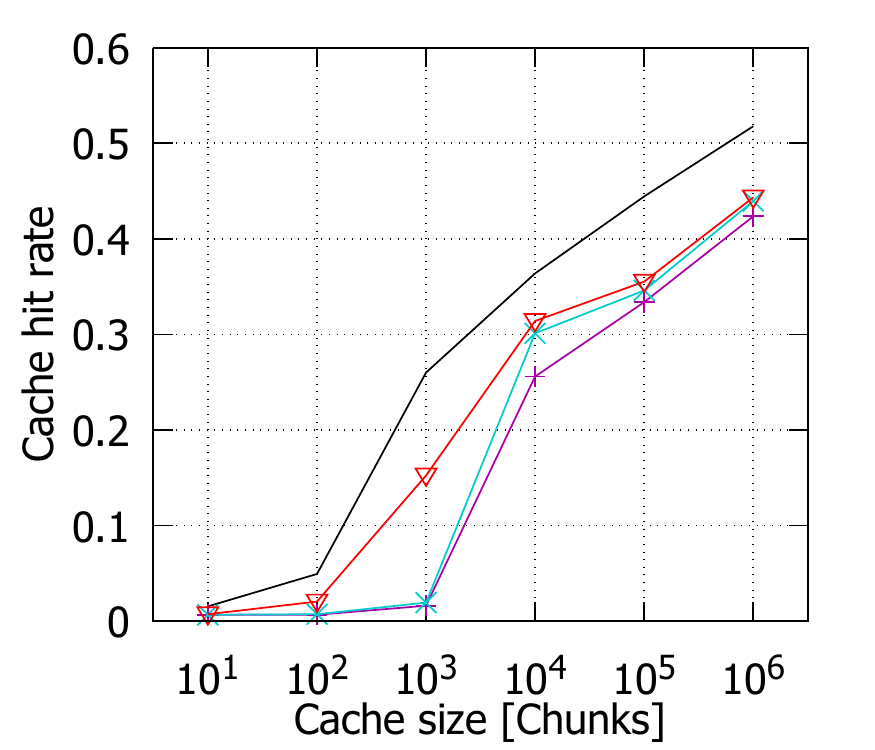}{\acsYPc}{car:acsypc}
\subgraphcachehitrate{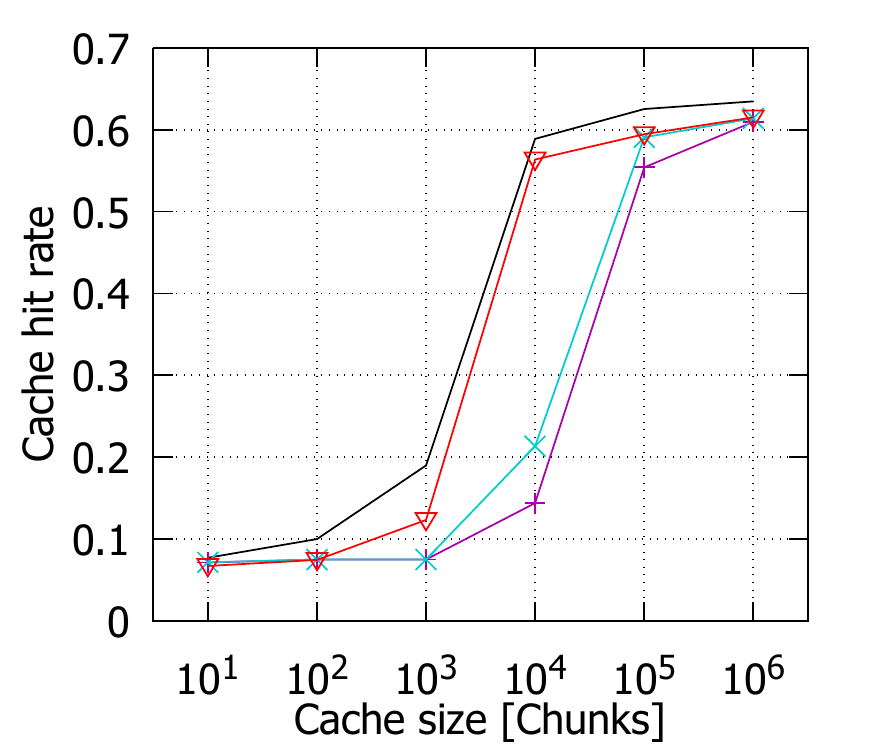}{\acsYPb}{car:acsypb}
\subgraphcachehitrate{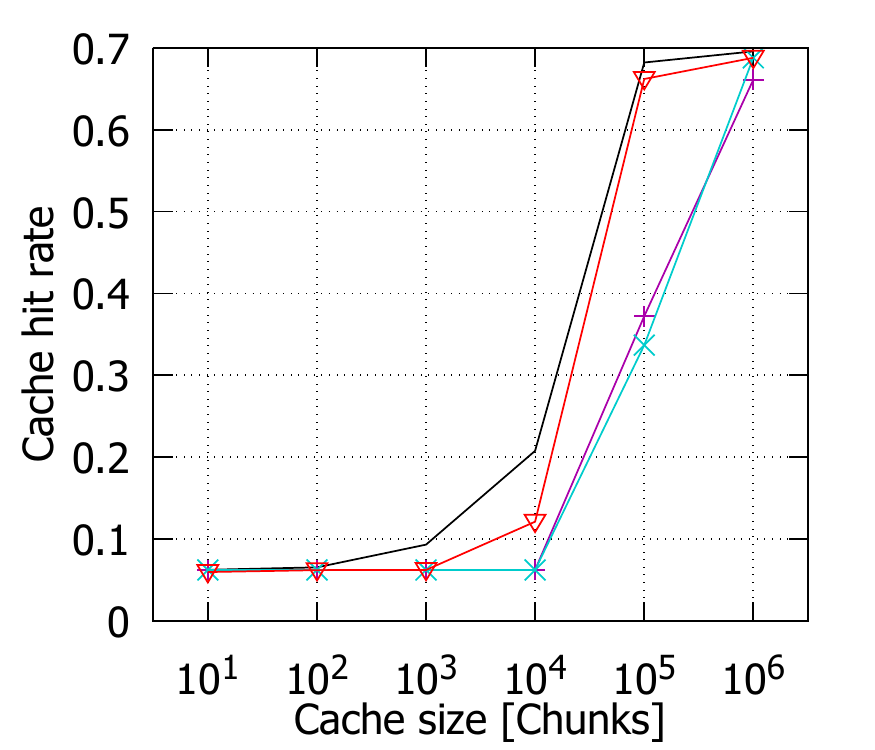}{\acsYPa}{car:acsypa}
\caption{Results for Real Traces}
\label{cache:hit:rate:car:acsy}
\end{figure*}

\begin{figure*}[t!]
\subgraphreusedistance{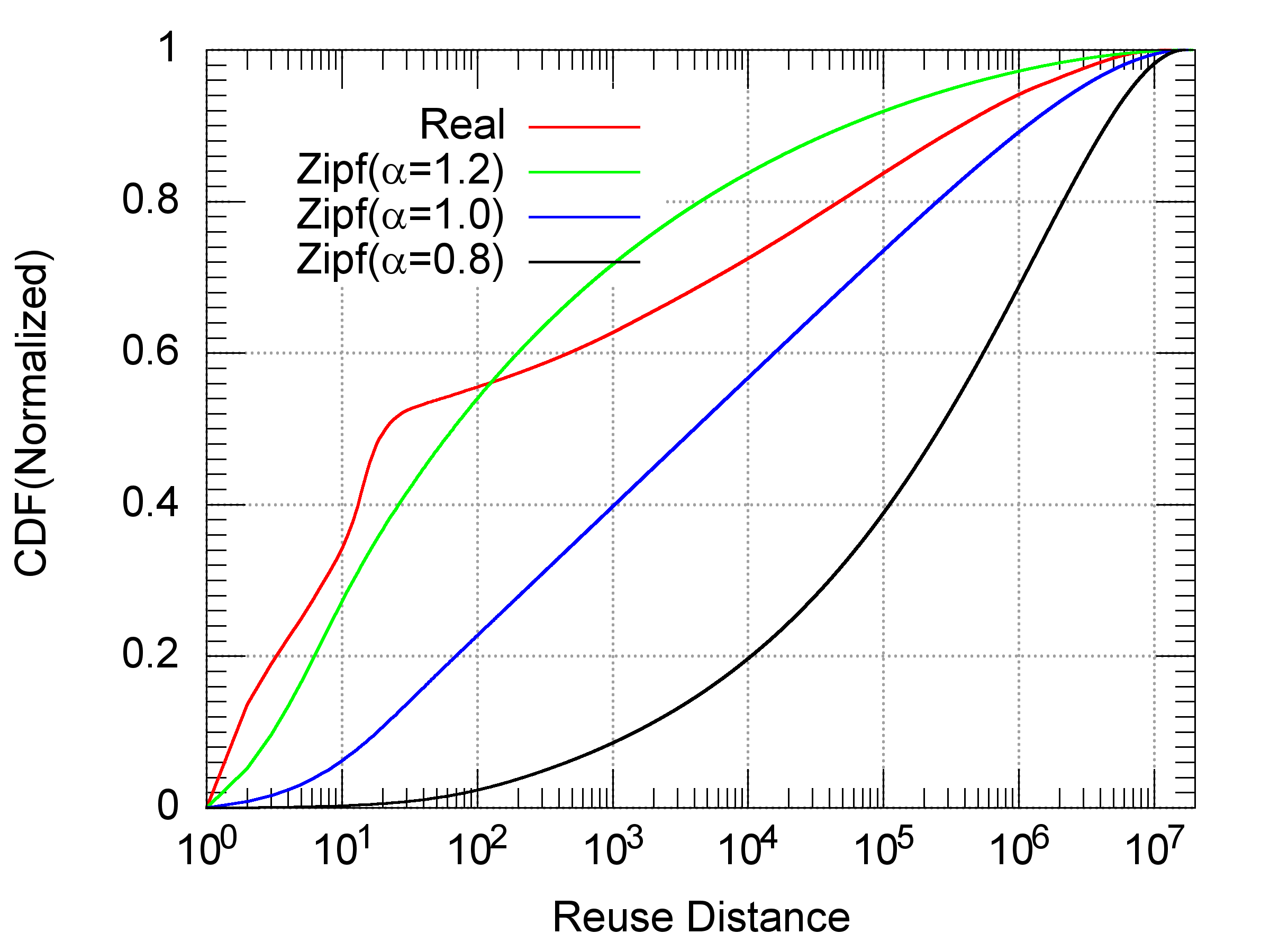}{Different Types of Workloads in Units of Content}{c}
\subgraphreusedistance{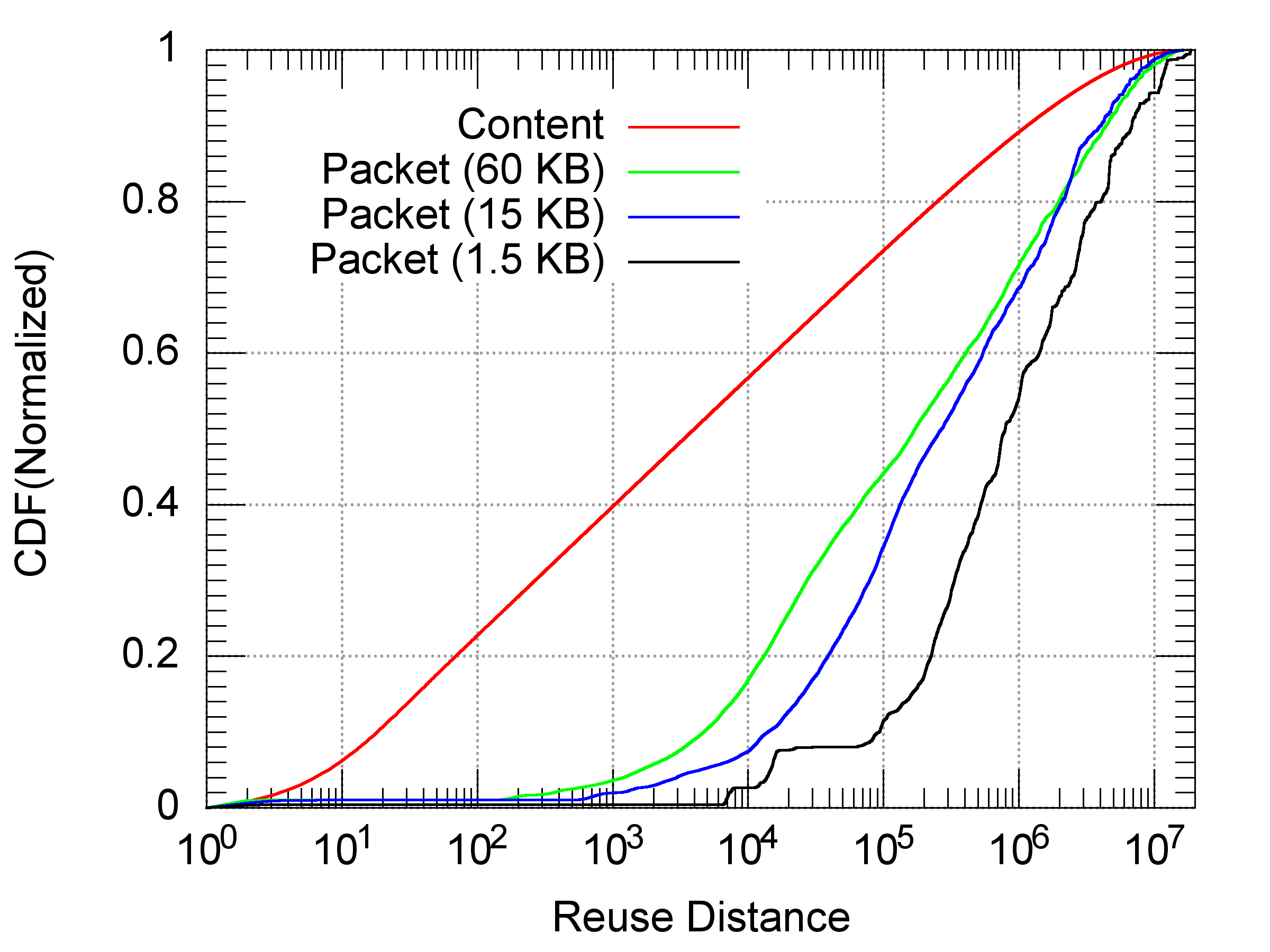}{Artificial Workloads ($\alpha=1.0$) in Different Units}{zcp}
\subgraphreusedistance{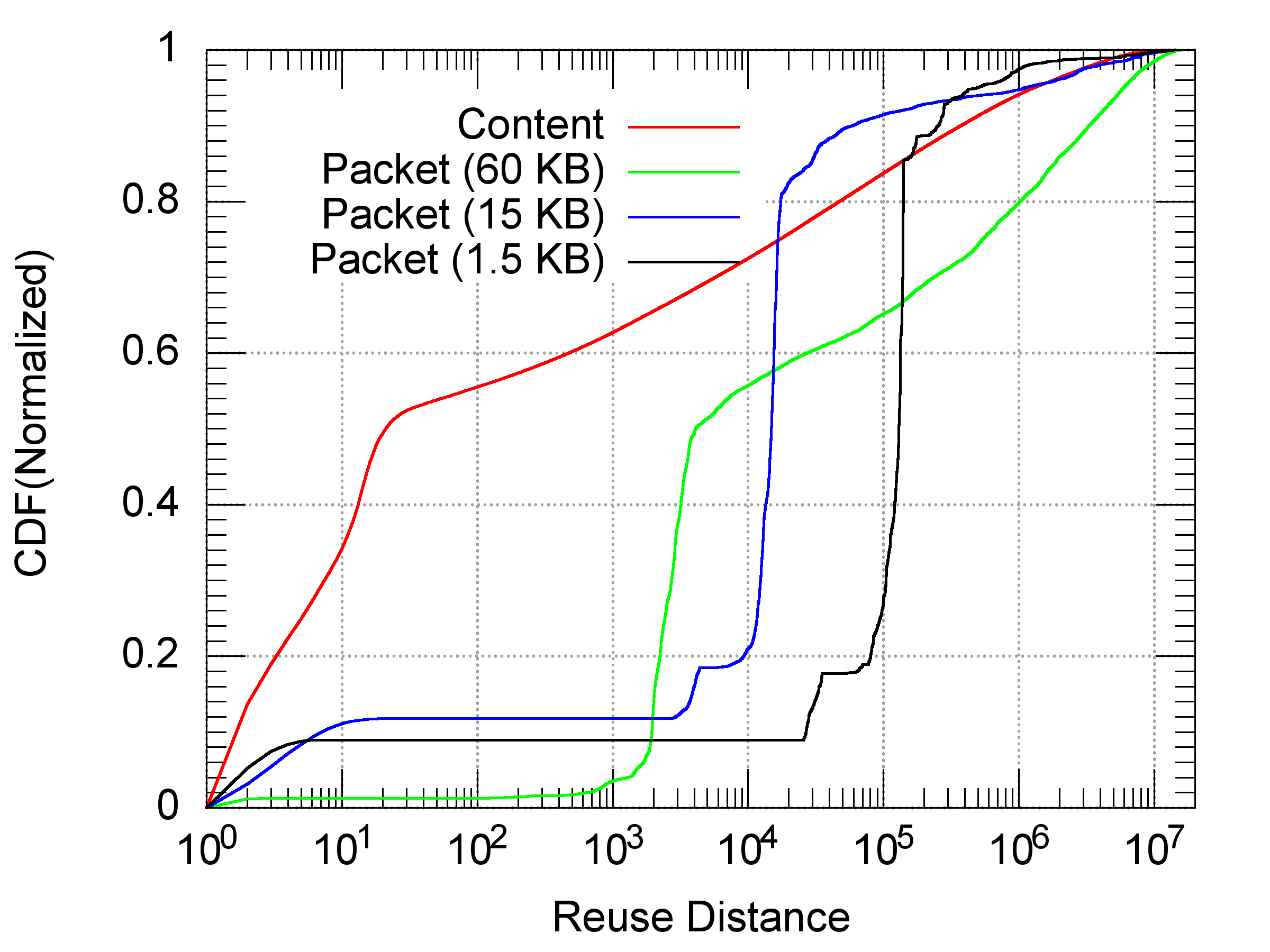}{Real Traces in Different Units}{ycp}
\caption{CDF of RD in Various Workloads}
\label{reuse:dist:acs}
\end{figure*}

\begin{figure*}[t!]
\subgraphcachehitratelinear{CAR_linear_z_080_c}{5}{\acsZbC}{car:acszbcl}
\subgraphcachehitratelinear{CAR_linear_z_100_p_60000}{3}{\acsZcPc}{car:acszcpclc}
\subgraphcachehitratelinear{CAR_linear_z_100_p_60000}{4}{\acsZcPc}{car:acszcpcld}
\subgraphcachehitratelinear{CAR_linear_a_p_15000}{3}{\acsYPb}{car:acsypbl}
\caption{Results for Simulation with a Linear Topology}
\label{cache:hit:rate:car:l}
\end{figure*}

\begin{figure*}[t!]
\subgraphimprovementratio{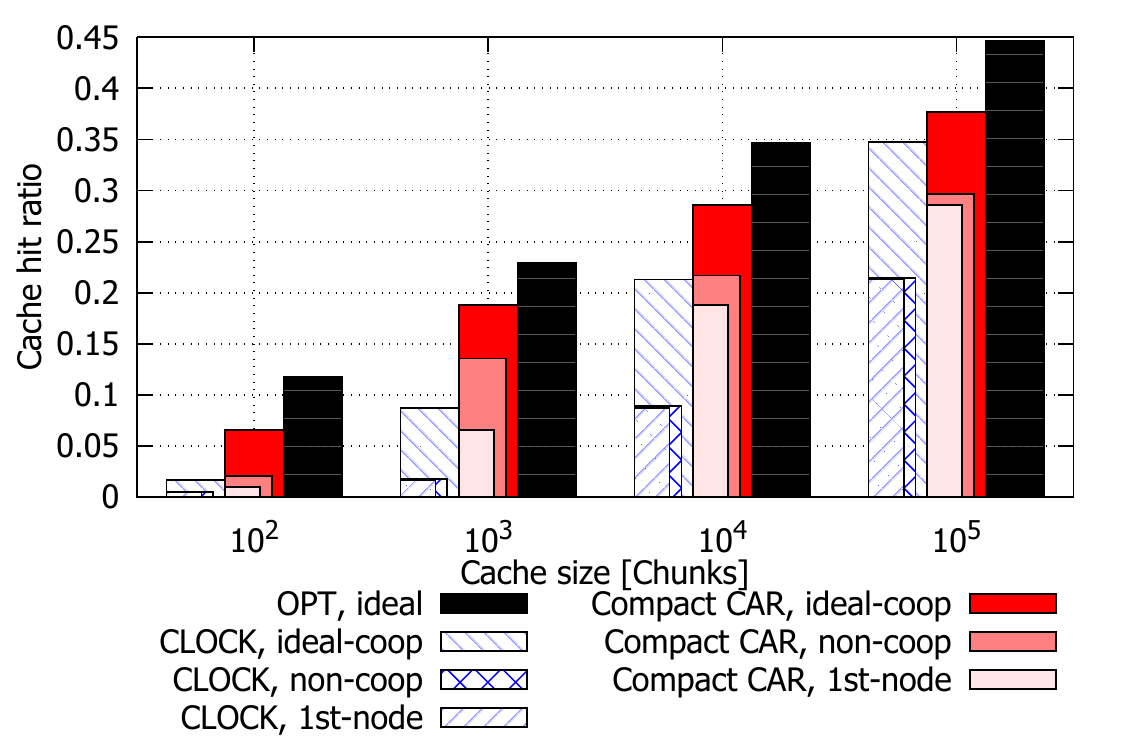}{\acsZcPb}{car:acszcpb}
\subgraphimprovementratio{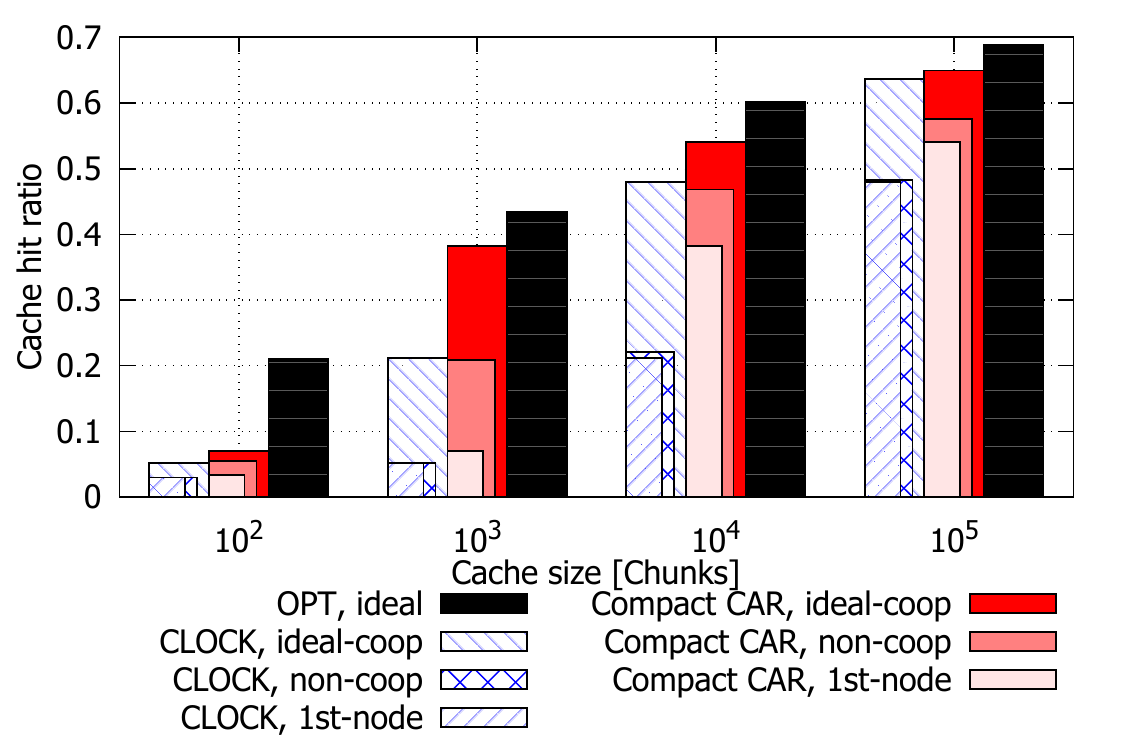}{\acsZdPc}{car:acsydpc}
\subgraphimprovementratio{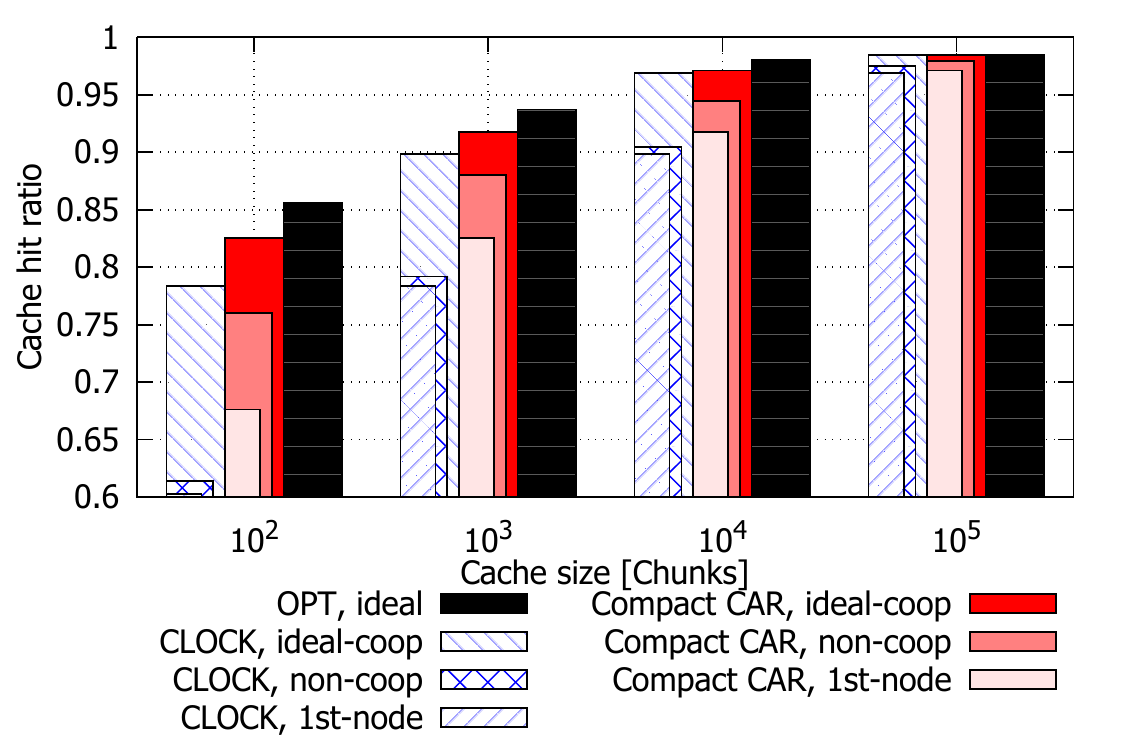}{\acsYC}{car:acsyc}
\caption{Comparison between non-cooperative caching and ideally-cooperative caching }
\label{cache:hit:rate:improve}
\end{figure*}

\begin{figure}[t!]
\begin{center}
\includegraphics[width = 7cm,clip]{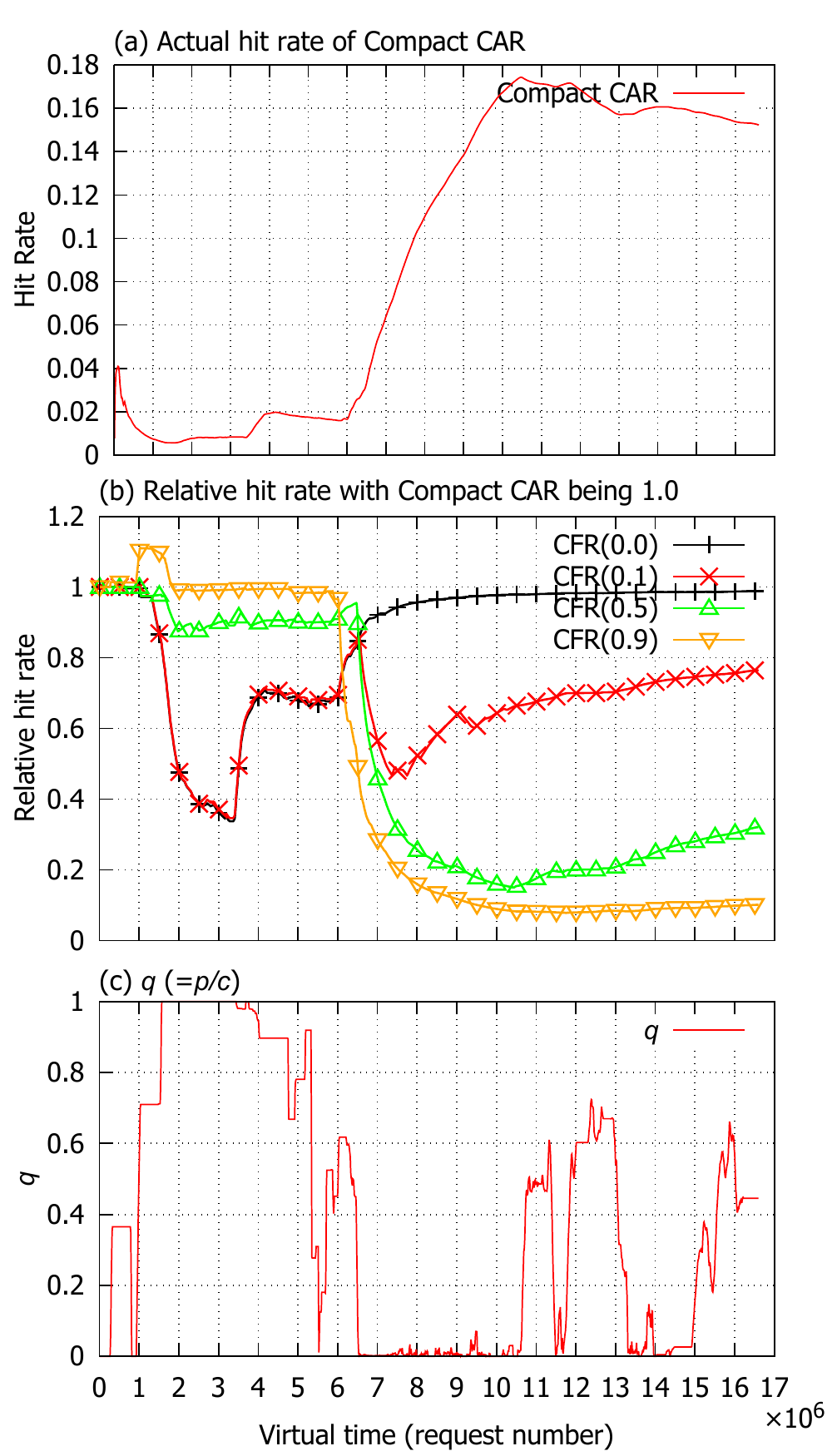}
\caption{Dynamics of Hit Rate of \cfr and Adaptive Parameter $q$}
\label{param:dynamics:rht:q}
\end{center}
\end{figure}

% \begin{figure*}[t!]
% \subgraphcarvscfr[0.5]{z100c}{\acsZcC}{car:vs:cfr:acszcc}
% \subgraphcarvscfr[0.5]{z100p01500}{\acsZcPa}{car:vs:cfr:acszcpa}
% \subgraphcarvscfr[0.5]{yp15000}{\acsYPb}{car:vs:cfr:acsypb}
% \subgraphcarvscfr[0.5]{yp60000}{\acsYPc}{car:vs:cfr:acsypc}
% \caption{Relative Cache Hit Rates of Compact CAR and \cfr}
% \label{cache:hit:rate:car:vs:cfr}
% \end{figure*}

Figures \ref{cache:hit:rate:car:acszc} depicts the cache hit rates of each cache replacement policy in a single ICN element with synthetic traffic described previously: \acsZC{} changing $\alpha$ from $0.6$ to $1.2$. 
Our proposal, Compact CAR, achieves a hit rate comparable to that of CAR, which is contrary to our speculation. We conjectured that the operation mixing the order in the Compact CAR would degrade its performance. The result is promising because we can achieve the performance as good as CAR even with much less memory cost. The memory cost of Compact CAR including several others is theoretically analyzed in Section \ref{sec:eval:cmplx} in detail.
In addition, the results show that Compact CAR can achieve the same cache hit ratio with one-tenth of cache size compared to simple cache replacement algorithms such as FIFO and CLOCK in the best case. 

In addition to the simulation using traces in units of contents, Figures \ref{cache:hit:rate:car:acszp} shows the cases when the sizes of cacheable chunks change from 60 KB to 1.5 KB with the parameters of the Zipf distribution ($\alpha$) at 1.0 and 1.2, which are denoted, e.g., $A_{Zipf(0.6)}^{P(60KB)}$ to $A_{Zipf(0.6)}^{P(1.5KB)}$.
As the value of $\alpha$ increases, the hit rate increases. This means that a high popularity bias results in a high hit rate.
As depicted in Fig.\ref{cache:hit:rate:car:acszc:ctop}, we observe that the cache hit rate decreases substantially as the size of cacheable items becomes small, e.g., from a whole content to chunks.

%%%%%%%%%%%%%%%%%%%%%%%%%%%%%%%%%%%%%%%%%%%%%%%%%%%%%%%%%%%%%%%%%%%%%%%%%%%%%%%%%%%%%%%%%%%%%%%%%%%%%%%%%%%%%%
\subsection{Cache Hit Rate with Real Traffic Trace}\label{sec:eval:real}

Figure \ref{cache:hit:rate:car:acsy} presents the simulation results with real Video-on-demand (VoD) traffic which was collected at Osaka University. Each content is segmented into small size chunks to simulate the transmission of chunks in ICN networks. The cache hit rates in Fig. \ref{cache:hit:rate:car:acsy} are similar to those in Fig. \ref{cache:hit:rate:car:acszc} and Fig. \ref{cache:hit:rate:car:acszp}. In Fig. \ref{cache:hit:rate:car:acsy}, one interesting observation is that the cache hit rate of our proposed algorithm suddenly soars, e.g., when cache size is $10^4$ in Fig. \ref{cache:hit:rate:car:acsypb} compared to conventional cache replacement algorithms: the performance becomes outstanding. This phenomenon correlates to the Reuse Distance (RD); therefore, we discuss it below.

Figure \ref{reuse:dist:acs} plots the cumulative distribution functions (CDFs) of RD. RD represents the number of chunks between two consecutive same chunks. For example, consider what happens when the value of RD is larger than the size\footnote{Its unit is the number of chunks} of cache. When the first chunk in the two consecutive same chunks is cached, it has a high probability to be discarded from the cache. If this case keeps happening due to a large amount of one-time contents (e.g., SCAN and LOOP), only non-popular contents remain in the cache. This situation is called cache pollution that non-popular contents occupy whole cache causing low-cache hit rate. Thus, a cache hit almost occurs when RD is smaller than cache size, and vice versa.

As mentioned in Section \ref{sec:key:idea}, Compact CAR maintains two link lists: one for non-popular contents, and the other for popular contents. Thus, the cache pollution only affects to the link list that maintains non-popular contents. In other words, Compact CAR is robust to the cache pollution scenario caused by a large amount of non-popular contents.

%%%%%%%%%%%%%%%%%%%%%%%%%%%%%%%%%%%%%%%%%%%%%%%%%%%%%%%%%%%%%%%%%%%%%%%%%%%%%%%%%%%%%%%%%%%%%%%%%%%%%%%%%%%%%%
\subsection{Simulation with a Line Topology}\label{sec:eval:line}

 The results with a line topology are shown for the purpose of showing the lower bound of the performance where a cooperative caching mechanisms fail. As mentioned previously, the cache hit rate is governed by two factors: one is a cache replacement algorithm (how to cache), and the other is a cooperative caching algorithm (where to cache). We can assume that one node topology represents a case where a cooperative caching algorithm works ideally. If one node whose caching capacity is equivalent to the total $n$ nodes, the one node topology can be considered as the $n$-node topology that has an ideal cooperative caching mechanism. In other words, the result with one node topology shows the upper bound of the performance where cooperative caching works ideally. On the other hand, the line topology represents non-cooperative caching algorithms, especially when contents being downloaded from one end to the other are cached every nodes in the line topology. In this case, the cache capacity of a whole network is considerably wasted by redundant caches. Thus, our simulation results clarify the upper and lower bound of the performance caused by cooperative caching mechanisms.

Figure \ref{cache:hit:rate:car:l} presents the cache hit rates of individual nodes on a line topology. Compact CAR improves the hit rate in the second and succeeding routers, whereas the hit rate of FIFO and CLOCK decreased to approximately zero. Figure \ref{cache:hit:rate:improve} shows the upper and lower bound of the performance achieved by cooperative caching. We denote the performance of ideally cooperative caching by ``ideal-coop'', which specifies the upper bound. The result denoting ``non-coop'' means the total cache hit rates of nodes in a line topology, which is the performance of non-cooperative caching and specifies the lower bound. We also show the hit rate of the only first node of a line topology as ``1st-node'' to understand how CLOCK is inappropriate for the environment without cooperative caching. There is less difference between the upper bound and the lower bound of Compact CAR than that of CLOCK. This result indicates Compact CAR can exploit resources in a network by reducing redundant caches caused by the cooperation failure. 

It is interesting to analyze the performance under an environment with a certain cooperation or a cache decision algorithm; however, we do not show the analysis because the main purpose of this paper is proposing the cache replacement algorithm that is feasible and appropriate for an ICN element. In future, we will investigate the effects of various cache placement and decision algorithms on a network and communication quality. 

\subsection{Dynamic Parameter Tuning}\label{sec:eval:param:tuning}

As explained in Section \ref{sec:design}, Compact CAR dynamically adapts to changing traffic access patterns by varying the parameter $p$.
There is no one-size-fits-all parameter and it is necessary that the parameter should be tuned to maximize a cache hit rate under any circumstances.

Here we evaluate the parameter tuning strategy for the proposed Compact CAR whose parameter $p$ represents the target size of $T_1$. The parameter $p$ ranges from zero to the cache size $c$. As the value of $p$ increases, the operational behavior of Compact CAR becomes similar to the case where recently accessed content becomes important. On the other hand, as $p$ decreases, Compact CAR behaves similar to the case where frequently requested content becomes important.

Thus, depending on the variation of access patterns, the parameter $p$ should be tuned.
To compare the difference between dynamical tuning and statical tuning, we introduce Clock with Fixed Replacement (CFR) algorithm which corresponds to our proposal Compact Clock with Adaptive Replacement (Compact CAR). CFR has the fixed value of $q=p/c$ ($0\leq q \leq 1$) which is determined in advance.

Figure \ref{param:dynamics:rht:q} shows that Compact CAR adaptively changes the parameter: the trends of $q$ and the cache hit rates of CFR($q$). The $x$-axis shows the virtual time $t$, which is equivalent to the total number of requests. The cache hit rates of CFR($q$) are shown as relative value with that of Compact CAR being 1.0 in Fig. \ref{param:dynamics:rht:q} (b).
When $0<t<6\times 10^6$, the hit rate of CFR($q$) with high $q$ tends to increase as $q$ increases, and vice versa.
The results show that Compact CAR can adaptively change the parameter. When $t=6\times 10^6$, we can observe the rapid increase in the cache hit rate of Compact CAR.
This increase is due to an arrival of many popular contents. Thus, the value of $q$ decreases to adopt the access patterns, where frequently accessed content becomes important, and the corresponding hit rate of CFR($q$) increases.
In addition, $q$ of Compact CAR continues to follow the optimal value at any time as evidenced by the fact that the best relative hit rates among CFR($q$) are at most nearly 1.0. By contrast, the relative hit rates of the parameter fixed algorithms become at worst nearly 0.1. Thus, we can confirm that the parameter tuning algorithm of Compact CAR are necessary and greatly adaptive.

%%%%%%%%%%%%%%%%%%%%%%%%%%%%%%%%%%%%%%%%%%%%%%%%%%%%%%%%%%%%%%%%%%%%%%%%%%%%%%%%%%%%%%%%%%%%%%%%%%%%%%%%%%%%%%
\subsection{Analysis on Space and Time Complexities of CAR and Compact CAR}\label{sec:eval:cmplx}

\begin{table*}[t!]
\begin{center}
\caption{Time Complexity of Cache Replacement Algorithm's Overhead}
\label{cache:rpl:time:complexity}
\footnotesize
\begin{tabular}{c|cc|cc}
\hline
         & \multicolumn{2}{c|}{worst case} & \multicolumn{2}{|c}{average case}  \\
policies & hit & miss                      & hit & miss                          \\
\hline
FIFO                       &  $\delta$           & $t_r+t_w+\delta$    & $\delta$           & $t_r+t_w+\delta$          \\
\lrudll                    &  $3t_r+6t_w+\delta$ & $3t_r+6t_w+\delta$  & $3t_r+6t_w+\delta$ & $3t_r+6t_w+\delta$        \\
\lrus                      &  \oN                & \oN                 & \oN                & \oN                       \\
\lruc                      &  \oc                & \oN                 & \oc                & \oN                       \\
\lfuh                      &  \ol                & \ol                 & \ol                & \ol                       \\
ARC (with \lrudll)         &  \oc                & \oc                 & \oc                & \oc                       \\
LIRS (with \lrudll)        &  \oM                & \oM                 & $O(\frac{1}{\beta})$   & $O(\frac{1}{\beta})$          \\
CLOCK                      &  $t_w+\delta$       & \oN                 & $t_w+\delta$       & $O(\frac{1}{1-\beta})$      \\
CAR (with \lrudll)         &  $t_w+\delta$       & \oN                 & $t_w+\delta$       & $O(\frac{1}{1-\beta})$    \\
% CLOCK-Pro                  &  $t_w+\delta$       & \oN                 & $t_w+\delta$       & $O(\frac{1}{(1-\beta)^2})$  \\
Compact CAR (our proposal) &  $t_w+\delta$       & \oN                 & $t_w+\delta$       & $O(\frac{1}{1-\beta})$    \\
\hline
\end{tabular}
\end{center}
\begin{center}
\caption{Space Complexity of Cache Replacement Algorithm's Overhead}
\label{cache:rpl:space:complexity}
\footnotesize
\begin{tabular}{c|cc|c}
\hline
         & \multicolumn{2}{c|}{Space Complexity} &                \\
policies &  memory [bit]  & order                 & number of history \\
\hline
FIFO                       & $\log{n}$                         & \ol             & -   \\
\lrudll                    & $2n\log{n}+2\log{n}$              & \oNl            & -   \\
\lrus                      & $\delta$                          & \oc             & -   \\
\lruc                      & $n\log{n}+\log{n}$                & \oNl            & -   \\
\lfuh                      & $n\cdot{}C$                       & $O(n\cdot{}C)$  & -   \\
ARC (with DLL)             & $4n\log{n}+7\log{n}$              & \oNl            & $n$ \\
LIRS (with DLL)            & $4n\log{n}+2n+2m\log{n}+4\log{n}$ & $O(m+n\log{n})$ & $m$ \\
CLOCK                      & $n+\log{n}$                       & \oN             & -   \\
CAR (with DLL)             & $4n\log{n}+n+9\log{n}$            & \oNl            & $n$ \\
% CLOCK-Pro                  & $8n+6\log{n}$                     & \oN             & $n$ \\
Compact CAR (our proposal) & $n+9\log{n}$                      & \oN             & $n$ \\
\hline
\end{tabular}
\end{center}
\end{table*}

We analyze the time and space complexities of Compact CAR. The complexity is analyzed from the viewpoint of an additional process or memory required for the algorithms. 
In the evaluation of time complexity, we calculate the number of memory access as a dominant factor when a cache hit or a miss occurs. Because the actual value is typically unsteady, we study the worst-case and average-case complexity in the two different cases (i.e., a cache hit and a cache miss). Space complexity depends on the amount of additional bits needed to maintain a data structure, and so we calculate the amount of bits. We also express them with big $O$ notation. Our analysis does not calculate the amount of memory to keep ghost caches since it should be compared with the amount of memory required for cache data rather than control information.

In this analysis, we define the following notations and variables. 
$n$ is the number of cache entries.
Some policies use $P$-bit pointers to cache entries.
$P$ requires at least $\lceil \log{n} \rceil$ [bit] to identify $n$ individual entries. 
For the analysis of the time complexities of variants of CLOCK, let us assume $h_i$ denotes the number of content accessed at least $i$ times in a certain range, $\beta$ and $\gamma$ represent $h_2/h_1$ and $h_3/h_1$, respectively. Note that $\beta$ and $\gamma$ satisfies the inequality $0\leq \gamma \leq \beta \leq 1$ since $h_{i+1} \leq h_{i}$. 
We basically express time complexity of an algorithm as order of the function of $n$ or $\beta$. If the complexity of a algorithm is $O(1)$ and can be accurately calculated, we describe the complexity with read time $t_r$, write time $t_w$ and negligibly small time $\delta$, which is required for the other processes, instead of big $O$ notation, because the memory access time is a dominant factor in caching algorithm execution time.

Although we analyze only two cache replacement algorithms: CAR and Compact CAR, Table. \ref{cache:rpl:space:complexity} and \ref{cache:rpl:space:complexity} summarize the analytical results of space and time complexities of not only the two of them but also other cache replacement algorithms including FIFO, LRU, CLOCK, ARC and LIRS for the purpose of comparison. The detail explanations on the complexity analysis for other than CAR and Compact CAR are presented in the \ref{sec:cmplx:others}.

\subsubsection{Space Complexity}\label{sec:eval:cmplx:space}

First, we analyze the space complexity of CAR and Compact CAR. Compact CAR maintains four CLOCKS shown in Fig. \ref{arclock:shift:border:a}. Our simple swapping renders Compact CAR free from the additional costs of memory or process for maintaining the order of sweeping the list.
Furthermore, $B_1$ and $B_2$ do not need $R$-bit and the total length of the other two CLOCKs, $T_1$ and $T_2$, is $n$. Thus, Compact CAR costs $(n+9P)$ bits for two normal CLOCKs whose total length is $n$, two CLOCKs without a $R$-bit, four information of the size of the lists, and a parameter of a target size.

On the other hand, CAR has two variable-size CLOCK lists and two LRU lists. The variable-size CLOCK must support insertion (deletion) of a chunk into (from) an arbitrary position in a list allocated in physically contiguous memory. The implementation of variable-sized CLOCK needs the same data structure as LRU to keep the order of sweeping the list. Because the approach illustrated in Fig. \ref{cost:insertion:car}(b) imposes no additional memory cost, we focus on CAR implemented with a doubly-linked list.
Space complexity of two CLOCK lists and two LRU lists is comparable to that of four doubly-linked lists whose total maximum length is $2n$. In addition, total $n$ R-bits are required for two CLOCK lists. CAR also uses an adaptively tuned parameter called a target size, which costs at least $P$ bits. Thus, the memory overhead is $(4Pn+n+9P)$ bits. 

\subsubsection{Time Complexity}\label{sec:eval:time}

Second, we elaborate the time complexity of them. Since many of the analysis is overlapped, we first elaborate the time complexity of CAR, followed by that of Compact CAR.
CAR as well as CLOCK incurs $t_w+\delta$ complexity at a cache hit since it requires only to update $R$-bit. The worst-case complexity at a cache miss is $O(n)$ because a hand must move $n$ times to go around the clock in the worst case where R-bit of all entries in CLOCK is set. 

The average number of hand movements at a cache miss $\omega$ is represented as $n/s$, where $s$ is the number of cache misses during $n$ hand movements. Because we aim to calculate the order of $\omega$, our analysis can be simplified by considering the extreme case where $\omega$ is maximized in the steady state. Therefore, we consider two cases where $n$ is maximized, and where $s$ is minimized. For brevity, we do not show how to maximize $n$ and minimize $s$ here, which is obtained by the same calculation as CLOCK discussed in \ref{sec:cmplx:clock}. The difference between CLOCK and CAR is that we must count not only the first and second accesses to a chunk but also the third accesses should to maximize $n$ since the accesses turn on $R$-bits of entries in $L_2$. According to the calculation, $\omega$ satisfies the following inequality:
\[
\omega =\frac{n}{s}\leq
\frac{h_1+h_2+h_3}{h_1-h_2}=\frac{1+\frac{h_2}{h_1}+\frac{h_3}{h_1}}{1-\frac{h_2}{h_1}}=\frac{1+\beta+\gamma}{1-\beta}.
\]
Thus, the average-case time complexity depends on the characteristics of accesses rather than the cache size $n$, and $O(\omega)=O(\frac{1+\beta+\gamma}{1-\beta})=O(\frac{1}{1-\beta})$ because $0\leq \gamma \leq zeta \leq 1$. 
Time complexity of Compact CAR can be calculated in the same way as CAR. Time complexity at a cache hit is $t_w+\delta$, and worst-case and average-case complexity at a cache miss are $O(n)$ and $O(\frac{1}{1-\beta})$, respectively.

%%%%%%%%%%%%%%%%%%%%%%%%%%%%%%%%%%%%%%%%%%%%%%%%%%%%%%%%%%%%%%%%%%%%%%%%%%%%%%%%%%%%%%%%%%%%%%%%%%%%%%%%%%%%%%
\section{Discussion on the Implementation of Compact CAR for High Performance ICN Core Element}\label{sec:discuss}

%%%%%%%%%%%%%%%%%%%%%%%%%%%%%%%%%%%%%%%%%%%%%%%%%%%%%%%%%%%%%%%%%%%%%%%%%%%%%%%%%%%%%%%%%%%%%%%%%%%%%%%%%%%%%%
\subsection{Computational Overhead of Variants of CLOCK}\label{sec:discuss:computation}

In the previous section, we analyzed the computational cost of Compact CAR, which provides the complexity of $O(1/(1-\beta))$ in terms of $\beta$. It may be arguable that the complexity could be extremely large as the parameter $\beta$ becomes close to one.
In fact, the $\beta$ values of content-level and packet-level workloads used in our simulation ranges from 0.38 to 0.71 and from 0.08 to 0.22, respectively. $\frac{1+2\beta}{1-\beta}$ showing average-case time complexity of Compact CAR is less than only $2.0$ when $\beta<0.2$. $\frac{1+2\beta}{1-\beta}$ grows $6.0$, which is the computational cost of LRU, when $\beta$ becomes 0.625. $\frac{1+2\beta}{1-\beta}<8.0$ even if $\beta<0.7$. Although space complexity of CAR can be reduced by using a stack instead of a doubly-linked list, the implementation with a stack makes time complexity prohibitive as illustrated in Fig. \ref{cost:insertion:car}(b). 
In conclusion, the computational and memory costs of Compact CAR are acceptable in the design of high performance ICN core element.

%%%%%%%%%%%%%%%%%%%%%%%%%%%%%%%%%%%%%%%%%%%%%%%%%%%%%%%%%%%%%%%%%%%%%%%%%%%%%%%%%%%%%%%%%%%%%%%%%%%%%%%%%%%%%%
\subsection{Feasibility of Hardware Implementation}\label{sec:discuss:memory}

\begin{figure}[t!]
\begin{center}
\includegraphics[width=\columnwidth ,clip]{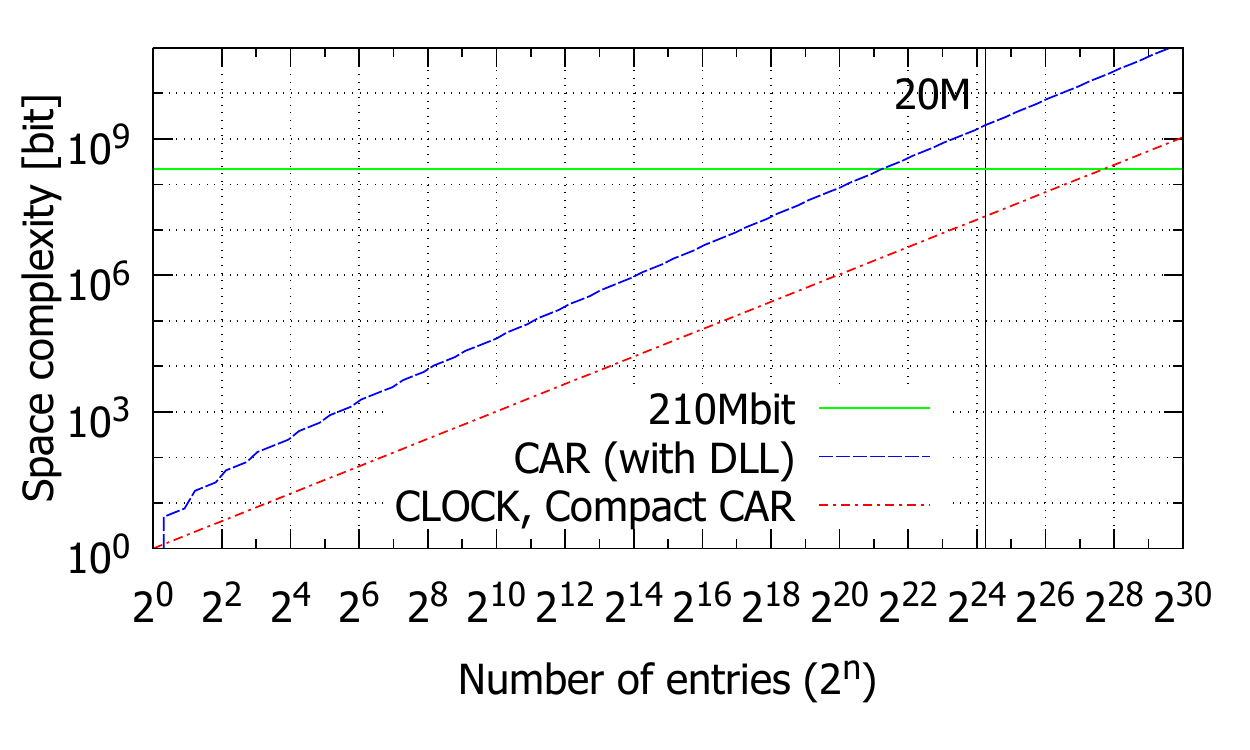}
\end{center}
\caption{Space Complexities of CAR and Our Proposal(Compact CAR)}
\label{spacel:cmplx:car}
\end{figure}

The throughput and capacity of a cache are the most serious obstacles to realize in-network caching. Assuming 10 Gbps of traffic and with 64-byte data packets, a single-line card would not to have throughput of approximately 20 million accesses per second at maximum (equivalently, 50 ns access time at a minimum). Since routers typically contain many line cards, a cache mechanism in a router must realize a level of throughput in linear proportion to the number of line cards. In practice, the existence of interest packets, data packets larger than 64 bytes, and skipping cache accesses by cache hit may ease the required access time several-fold; however, a cache decision policy is necessary to enable ten-fold improvement. 

Figure \ref{spacel:cmplx:car} shows a memory overhead of CAR and Compact CAR. 
As explained in \ref{sec:eval:cmplx:space}, CAR using a doubly-linked list consumes $(4Pn+n+9P)$ bits. Assuming a router holds 20 million cache entries, the memory cost of CAR becomes 2 Gbit to hold 20 million entries because $P\geq \lceil \log{n} \rceil$. This cost is prohibitive according to the constraint of SRAM, whose available size is 210 Mbit~\cite{ccn001}. 
On the other hand, Compact CAR requires a memory overhead of one bit per chunk. Compact CAR consumes 20 Mbit; therefore the cost of Compact CAR is feasible.

If SRAM access time is 0.45 ns~\cite{ccn001}, the router can handle the traffic of the four line cards quickly enough to keep up, even in the case of a cache miss causing several rotations of the hand and several swapping processes. However, the data of the chunks must be kept in a scalable memory, such as dynamic RAM or a solid-state disk. Since such memory is slow, we plan to consider a hierarchically structured cache memory and a pipelined process to ensure a high average speed for read/write accesses. We will eventually evaluate the router performance in a hardware implementation of the router, combining Compact CAR and a name lookup entity~\cite{a-ooka15ieice-HighSpeedRouter}, to demonstrate the feasibility of the router.

%%%%%%%%%%%%%%%%%%%%%%%%%%%%%%%%%%%%%%%%%%%%%%%%%%%%%%%%%%%%%%%%%%%%%%%%%%%%%%%%%%%%%%%%%%%%%%%%%%%%%%%%%%%%%%
\section{Conclusions}\label{sec:conclusion}

Few researches have been done for cache replacement algorithms in the context of ICN because they have been intensively researched in the fields of web-caching and content delivery network previously.
This paper argued that the conventional cache replacement algorithms cannot be directly applied to the design of a high performance ICN core element.

For this reason, we proposed a novel cache replacement algorithm named Compact CAR which would be an important component in the design of a high performance ICN core element.
Compact CAR outperforms compared to conventional cache replacement algorithms in terms of cache hit rates and reduction of memory usage in the design of ICN element.
In detail, the proposed algorithm can achieve the same cache hit rates with only one-tenth of memory usages that simple conventional algorithms use.
In addition, the cache hit rate by the proposed algorithm is only 10\% less than the optimal case over the various simulation scenarios. In particular, the difference becomes negligible when we use real traffic traces whose RD values are similar to the cache size.
This result provides a clue that a high cache hit rate can be achieved if the cache size adaptively changes according to the distribution of RD value in real traffic.
Furthermore, Compact CAR can dynamically adapt itself to the network environment whose traffic access patterns change dynamically, which is important to deal with various traffics in ICN.

ICN has been researched nearly 10 years and it may be the time to consider its deployment issue in Internet-Scale where the design of a high ICN core element becomes critical. We believe that the proposed cache replacement algorithm plays a key role in the design of such a high performance ICN core element in near future.

\section*{Acknowledgment}

This work was supported by the Strategic Information and Communications R\&D Promotion Programme (SCOPE) of the Ministry of Internal Affairs and Communications, Japan.

\bibliographystyle{elsarticle-num.bst}
\bibliography{\refdir/aooka_refs_IEEE,\refdir/a-ooka_e}

% \profile[\srcdir/profile/a-ooka.eps]{Atsushi Ooka}{received an M.E. degree from the Graduate School of Information Science and Technology, Osaka University, in 2014. Currently, he is a Ph.D. candidate. His research interests include the design and implementation of a router hardware for content-centric networking.}

% \profile[\srcdir/profile/ata.eps]{Shingo Ata}{received M.E. and D.E. degrees in Informatics and Mathematical Science from Osaka University in 1998 and 2000, respectively. Since 2013, he has been a professor at the Graduate School of Engineering at Osaka City University. His research work includes networking architecture, design of communication protocols, and performance modeling of communication networks. He is a member of IEICE, IEEE, and ACM.}

% \profile[\srcdir/profile/murata.eps]{Masayuki Murata}{received his M.E. and D.E. degrees from Osaka University, Japan, in 1984 and 1988, respectively. In 1984, he joined the Tokyo Research Laboratory, IBM Japan, as a researcher. He was an assistant professor from 1987 and an associate professor at Osaka University from 1992 to 1999. Since 1999, he has been a professor at Osaka University, and he is now with the Graduate School of Information Science and Technology, Osaka University. He has more than 500 papers in international and domestic journals and conferences. His research interests include computer communication networks, performance modeling, and evaluation. He is a member of IEICE, IEEE, and ACM.}

\clearpage

\appendix

\section{Time and Space Complexity of the Remaining Policies} \label{sec:cmplx:others}

We analyze time and space complexity of policies which are skipped in Section \ref{sec:eval:cmplx}.
The complexity is calculated based on an additional process or memory required for the algorithms. In the evaluation of time complexity, we calculate the number of memory access as a dominant factor when a cache hit or a miss occurs. Because the actual value is typically unsteady, we study the worst-case and average-case complexity in the two different cases (i.e., a cache hit and a cache miss). Space complexity depends on the amount of additional bits needed to maintain a data structure, therefore, we calculate the amount of bits. We also express them with big O notation. Our analysis does not calculate the amount of memory to keep ghost caches since it should be compared with the amount of memory required for cache data rather than control information. 

In addition to the notations in Section \ref{sec:eval:cmplx}, we define the following notations and variables. 
$m$ is the number of ghost cache entries in LIRS.
Statistical policies assign $C$-bit information (as a counter used in LFU) to every entry. 

\subsection{Complexity of FIFO}\label{sec:cmplx:fifo}

In FIFO, only a $P$-bit pointer to remember the head of the queue is required. When a cache hit
occurs, no additional operations are necessary (except for common operations such as reading the
accesses entry). When a cache miss occurs, there are two additional operations: reading the pointer
to evict the entry at the head of the queue and updating it. Thus, space complexity is $P$ bits.
Time complexity at a cache hit and miss is $\delta$ and $(t_r+t_w+\delta)$, respectively.

\subsection{Complexity of LRU}\label{sec:cmplx:lru}

\paragraph{\lrudll}

To implement \lrudll{}, it is necessary to maintain a sorted doubly-linked list, where each entry has two $P$-bit pointers and the most recently used (MRU) entry is at the front of the list. In addition, two pointers are needed to remember MRU and LRU entries. Thus, \lrudll{} totally requires $(2Pn+2P)$-bit memory overhead. 

Let us denote an entry by $e_i (i=1,2, \cdots, n)$, where smaller $i$ means that the entry is more recent, and its pointers that point previous and next entries by $p^{prev}_i$ and $p^{next}_i$, respectively. If $e_i$ is accessed, $e_i$ is moved to the front of the list. This process updates six pointers: two pointers of $e_i$, $p^{next}_{i-1}$, $p^{prev}_{i+1}$, $p^{prev}_{1}$ and a MRU pointer. To find $e_{i-1}, e_{i+1}$ and $e_{1}$, it is necessary to read three pointers. On the other hand, if there is a cache miss, $e_n$ is evicted and a new entry is cached as a previous entry of $e_1$. After reading the addresses of first, $n$-th and $(n-1)$-th entries, it is required to write a new entry and update $p^{next}_{n-1}$ and $p^{prev}_{1}$ and MRU and LRU pointers. Consequently, $(3t_r+6t_w+\delta)$ gives an estimate of time complexity imposed by \lrudll{} in the case of both a cache hit and a cache miss. 

\paragraph{\lrus}

\lrus{} introduces no additional memory cost because its data structure maintains all control information needed to perform the algorithm. LRU entry, which is evicted when a cache miss occurs, resides at the bottom of the stack. When a cache miss occurs, a new entry stored at the top of the stack.

However, \lrus{} requires shifting a large amount of entries to insert or move an entry just like the algorithm described in Section \ref{sec:design}.
If $e_i$ is accessed, all entries from $e_1$ to $e_{i-1}$ must be shifted. If there is a cache miss, it is required to shift entries from $e_1$ to $e_{n-1}$ and write a new entry at the top of the stack. In the worst case, $n$ entries are moved. On average, $n/2$ entries are moved at a cache hit if all entries are uniformly referenced. Thus, time complexity of \lrus{} is $O(n)$.
This process in a small-scale computer system is typically supported by special hardware for the shifting operation; however, it is infeasible for use in an ICN element because of an excessive amount of entries. 

\paragraph{\lruc}

\lruc{} assigns a $C$-bit counter to each entry. In addition, a $C$-bit counter is necessary to remember the total number of accesses. Thus, \lruc{} imposes $(Cn+C)$-bit space complexity. 

Time complexity at a cache hit is \oc in accordance with processes updating a counter and writing the value at a new entry. Time complexity at a cache miss is \oN{} because of the look-up process to retrieve an entry with the minimum counter value from the unsorted list.

\subsection{Complexity of CLOCK}\label{sec:cmplx:clock}

To store $n$ R-bits and a position located by a clock hand, space complexity of CLOCK is $(n+P)$ bits. 
Time complexity at a cache hit is $(t_w+\delta)$ since it requires only to update R-bit.
The worst-case time complexity at a cache miss is $O(n)$ because a hand must move $n$ times to go around the clock in the worst case where R-bit of all entries in CLOCK is set. However, such a case rarely happens. 

\begin{figure}[t!]
\begin{center}
\includegraphics[clip, width=\columnwidth ]{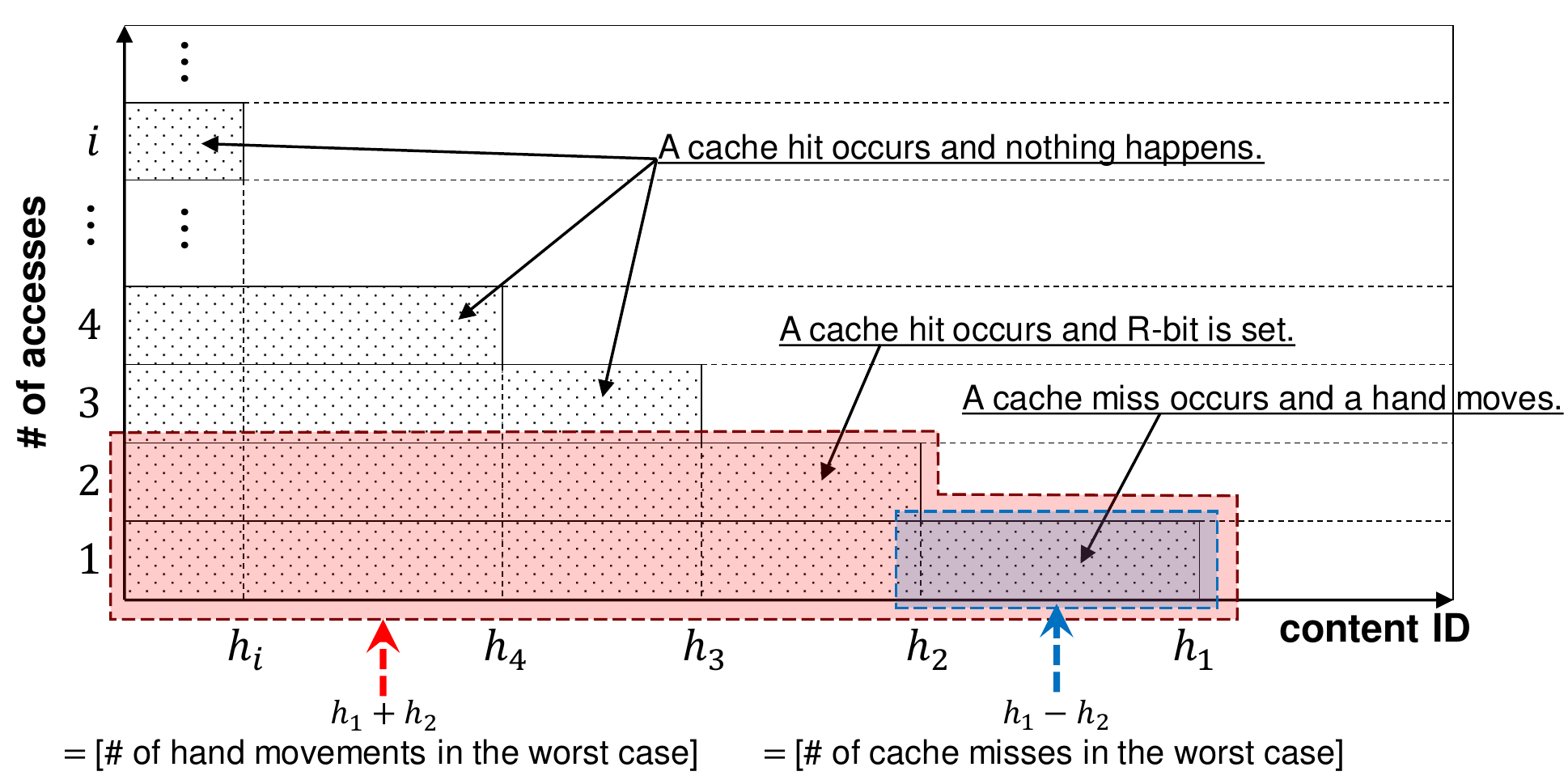}
\caption{Description of calculating $\omega$ of CLOCK}
\label{clock:complex:desc}
\end{center}
\end{figure}

Let $s$ denote the average number of cache misses during one cycle of a hand (i.e., $n$ hand movements) to calculate the average-case time complexity $\omega=n/s$, which can be defined as the number of hand movements per cache miss on average. Fig. \ref{clock:complex:desc} gives an intuitive understanding of how to calculate $n$ and $s$ according to $h_i$ defined in the time interval $[1,n]$ during $n$ hand movements. 

Because we aim to calculate the order of $\omega$, our analysis can be simplified by considering the extreme case where $\omega$ is maximized in the steady state. Therefore, we consider two cases where $n$ is maximized, and where $s$ is minimized. 

First, we discuss the case where $n$ is maximized. 
It is obvious that the first access to a chunk causes a cache miss and rotation of a hand. 
A cache hit by the second access to a chunk set R-bit of the accessed entry. 
This entry whose $R$-bit is set causes a movement of a hand because the hand ignores the entry only resetting the $R$-bit. 
Even if a chunk is accessed three or more times per cycle, the accesses do not cause a hand movement. 
Therefore, the number of hand movements to go around CLOCK's circular list is at most $h_1+h_2$ as illustrated in Fig. \ref{clock:complex:desc} (a red area). 

Second, we determine the minimum number of cache misses $s$. 
It is clear that $s=1$ at the minimum in the worst case where $(h_1-1)$ chunks have been already accessed and their $R$-bits are set before our considering time interval $[1,n]$. However, assuming the steady state where the popularity distribution of chunks (i.e. the distribution of $h_i$) is stable, there is at most $h_2$ chunks that is accessed before the beginning of the interval. Therefore, the number of cache misses is at least $h_1-h_2$ as illustrated in Fig. \ref{clock:complex:desc} (a blue area). 

According to the above discussion, $\omega$ satisfies the following inequality: 
\[
\omega =\frac{n}{s}\leq \frac{h_1+h_2}{h_1-h_2}=\frac{1+\beta}{1-\beta}.
\]
Thus, the average-case time complexity depends on the characteristics of accesses rather than the cache size $n$, and $O(\omega)=O(\frac{1+\beta}{1-\beta})=O(\frac{1}{1-\beta})$ because $0\leq \beta \leq 1$.

\subsection{Complexity of \lfuh}\label{sec:cmplx:lfu}

Because \lfuh is implemented with a heap, the complexity of \lfuh accords with that of a heap. If a heap is arranged in an array, $Cn$-bit space complexity is necessary because each entry holds a $C$-bit counter. The operation performed at a cache hit is moving an accessed entry, which is less expensive than adding a new entry. The operation performed at a cache miss is comparable to the cost of adding and deleting an entry. Both of the operations require \ol time complexity.

\subsection{Complexity of ARC}\label{sec:cmplx:arc}

ARC has two LRUs and each LRU contains $n$ entries, therefore, the space complexity of ARC implemented with \lrudll{} is more than twice as much as that of \lrudll{}. In addition, the LRU list is partitioned into two portions. To remember the partitioned location, each LRU list must maintain a $P$-bit pointer. ARC as well as CAR has the $P$-bit parameter. Thus, memory overhead of ARC grows $4Pn+7P$ bits. Time complexity is \oc as well as \lrudll{} because there is no repetition in ARC's algorithm.

\subsection{Complexity of LIRS}\label{sec:cmplx:lirs}

LIRS uses two LRUs which are called LRU stack $S$ and $Q$. The maximum size of LRU $S$ and $Q$ is $(n+m)$ and $n$, respectively. In addition, two bits are assigned to each entry to mark a hot chunk and a ghost cache. Thus, the space complexity is $(4Pn+2n+2Pm+4P)$. $m$ is practically smaller than $4n$~\cite{ccn116} although the length of $m$, which is determined by the length of a sequence of one-time content such as a scan and a loop, is theoretically unlimited.

Time complexity can grow significantly since there is an operation called stack pruning in LIRS. Stack pruning removes cold chunks that have not been accesses for a very long time including ghost caches. In the worst case, $m$ ghost caches are removed by only a single stack pruning operation, therefore, worst-case complexity is $O(m)$. Especially, if there is a long loop or scan, this overhead becomes extraordinarily large according to the length of the access pattern.

Average-case time complexity of stack pruning can be calculated in accordance with the average number of deleted entries by stack pruning, $\omega$. Assuming $n$ entries (i.e., the same amount of entries as the cache size) are removed by stack pruning while stack pruning is conducted $s$ times, $\omega$ can be defined as $n/s$. Specifying the time interval of $h_i$ accordingly, $h_1$ accesses causes cache misses, $h_2$ accesses render the accessed entry hot switching the LRU hot chunk into a cold chunk and trigger stack pruning. Because the other $\sum_{i\geq 3}{h_i}$ accesses treated as accesses to hot entries, stack pruning is not conducted by the accesses. According to the above calculations, the average-case time complexity is $O(\omega)=O(h_1/h_2)=O(1/\beta)$. The more one-time accesses occupy the traffic, the larger this complexity becomes.

\end{document}